%% file: main.tex
\documentclass{tcibook}
\usepackage{fancyhea}
\usepackage{work}
\usepackage{bm}       
\usepackage{graphicx}
\usepackage{multirow}
\usepackage{lineno}
\usepackage{threeparttable}
\usepackage[normalem]{ulem}
\usepackage{color}
\usepackage{amssymb}
\usepackage[shortlabels]{enumitem}
\usepackage[colorlinks = true,
            linkcolor = blue,
            urlcolor  = blue,
            citecolor = blue,
            anchorcolor = blue]{hyperref}

\input workshopsymbols.tex     
 
\usepackage{lineno}

\def\authorlist#1#2{
    \vskip 0.4in
\begin{center}\begin{large} {\bf  #1 } \end{large}
    \vskip 0.2in
              #2
     \vskip 0.2in
   \end{center}
}

\begin{document}

\pagenumbering{roman}

\parindent=0pt
\parskip=8pt
\setlength{\evensidemargin}{0pt}
\setlength{\oddsidemargin}{0pt}
\setlength{\marginparsep}{0.0in}
\setlength{\marginparwidth}{0.0in}
\marginparpush=0pt


\pagenumbering{arabic}

\renewcommand{\chapname}{chap:intro_}
\renewcommand{\chapterdir}{.}
\renewcommand{\arraystretch}{1.25}
\addtolength{\arraycolsep}{-3pt}
\renewcommand{\thesection}{\arabic{section}}

\input{RF6-report}

\bibliographystyle{JHEP}
\bibliography{RF6refs}

\end{document}

%% file: workshopsymbols.tex
\def\beq{\begin{equation}}
\def\eeq#1{\label{#1}\end{equation}}
\def\eeqn{\end{equation}}

\newenvironment{Eqnarray}%
   {\arraycolsep 0.14em\begin{eqnarray}}{\end{eqnarray}}
\def\beqa{\begin{Eqnarray}}
\def\eeqa#1{\label{#1}\end{Eqnarray}}
\def\eeqan{\end{Eqnarray}}

\let\bar=\overbar

\def\lsim{\mathrel{\raise.3ex\hbox{$<$\kern-.75em\lower1ex\hbox{$\sim$}}}}
\def\gsim{\mathrel{\raise.3ex\hbox{$>$\kern-.75em\lower1ex\hbox{$\sim$}}}}

\def\del{\partial}
\def\Dslash{\not{\hbox{\kern-4pt $D$}}}
\def\dslash{\not{\hbox{\kern-2pt $\del$}}}
\def\pslash{\not{\hbox{\kern-2pt $p$}}}
\def\ETmiss{\not{\hbox{\kern-4pt $E$}}_T}

\def\Dlr{\mathrel{\raise1.5ex\hbox{$\leftrightarrow$\kern-1em\lower1.5ex\hbox{$D$}}}}

\def\MSB{{\bar{M \kern -2pt S}}}
\def\msb{{\bar{\scriptsize M \kern -1pt S}}}

\def\drb{{\bar{\scriptsize D \kern -1pt R}}}



%% file: RF6-report.tex
\setcounter{chapter}{0} 

\chapter{Dark Sector Physics at High-Intensity Experiments}

\vspace{-2.5em}
\authorlist{Stefania Gori$^1$ and Mike Williams$^2$}
   {}
\vspace{-4em}
\begin{center}
\textbf{\underline{Section Editors} \\ \vspace{0.2em} Sec.~\ref{rf6-sec:experiments}: Phil Ilten,$^3$ Nhan Tran$^4$ \\ \vspace{0.2em}  Sec.~\ref{rf6-sec:BI1}: Gordan Krnjaic,$^{4,5}$ Natalia Toro$^6$  \\ \vspace{0.2em}  Sec.~\ref{rf6-sec:BI2}: Brian Batell,$^7$ Nikita Blinov,$^8$ Christopher Hearty,$^{9,10}$ Robert McGehee$^{11}$  \\ \vspace{0.2em}  Sec.~\ref{rf6-sec:BI3}: Philip Harris,$^2$ Philip Schuster,$^6$ Jure Zupan$^3$}\\ \vspace{1em}
\textit{
$^1$University of California, Santa Cruz, CA 95064, USA \\
$^2$Massachusetts Institute of Technology, Cambridge, MA 02139, USA\\
$^3$Department of Physics, University of Cincinnati, Cincinnati, OH 45221, USA\\
$^4$Fermi National Accelerator Laboratory, Batavia, IL 60510, USA \\
$^5$Department of Astronomy and Astrophysics, Kavli Institute for Cosmological Physics, \\ University of Chicago, Chicago, IL 60637, USA \\
$^6$SLAC National Accelerator Laboratory, Menlo Park, CA 94025, USA \\
$^7$Pittsburgh Particle Physics, Astrophysics, and Cosmology Center, Department of Physics and Astronomy, University of Pittsburgh, Pittsburgh, PA, 15260, USA \\
$^8$University of Victoria, Victoria, BC V8P 5C2, Canada \\
$^9$Department of Physics and Astronomy, University of British Columbia (UBC), \\ Vancouver, British Columbia, V6T 1Z1 Canada \\
$^{10}$Institute of Particle Physics (Canada), Victoria, British Columbia V8W 2Y2, Canada \\
$^{11}$Leinweber Center for Theoretical Physics, Department of Physics, \\
University of Michigan, Ann Arbor, MI 48109, USA
}
\end{center}

\vspace{-0.5em}

\textit{Is Dark Matter part of a Dark Sector?} The possibility of a dark sector neutral under Standard Model~(SM) forces furnishes an attractive explanation for the existence of Dark Matter (DM), and is a compelling new-physics direction to explore in its own right, with potential relevance to fundamental questions as varied as neutrino masses, the hierarchy problem, and the Universe's matter-antimatter asymmetry.  Because dark sectors are generically weakly coupled to ordinary matter, and because they can naturally have MeV-to-GeV masses and respect the symmetries of the SM, they are only mildly constrained by high-energy collider data and precision atomic measurements. Yet upcoming and proposed intensity-frontier experiments will offer an unprecedented window into the physics of dark sectors,  
highlighted as a \textit{Priority Research Direction} in the 2018 Dark Matter New Initiatives (DMNI) BRN report~\cite{BRN}.
Support for this program---in the form of dark-sector analyses at multi-purpose experiments, realization of the  intensity-frontier experiments receiving DMNI funds, 
an expansion of DMNI support to explore the full breadth of DM and visible final-state signatures (especially long-lived particles) called for in \cite{BRN}, 
and support for a robust dark-sector theory effort---will enable comprehensive exploration of low-mass thermal DM milestones, and greatly enhance  
the potential of intensity-frontier experiments to discover 
dark-sector particles decaying back to SM particles.

\textbf{The existence of dark matter motivates a dark sector.}   The existence of DM is firmly established due to its gravitational interactions. However, little is known about the particle nature of DM. A well-motivated possibility is that DM belongs to a dark sector of particles similar in structure (and possibly complexity) to that of ordinary matter. In contrast to the traditional weakly interacting massive particle (WIMP) paradigm, this dark sector is posited to be neutral under the SM forces. Similarly, all SM particles are neutral under the dark-sector forces. Therefore, a dark sector inherently explains the lack of strong or electromagnetic interactions of DM. Furthermore, DM charged under a new gauge interaction is inherently stable due to charge conservation in the dark sector, hence explaining the cosmological-scale DM lifetime. DM self-interactions mediated by dark-sector gauge bosons could also affect the dynamics of galactic structure formation, providing a natural explanation for some anomalous observations.  

\textbf{Dark sectors are a compelling possibility for new physics, to which intensity-frontier experiments offer unique and unprecedented access.}  In addition to the DM motivation, dark sectors naturally arise in many theoretically well-motivated models beyond the SM (BSM). For example, dark sectors that contain sterile neutrinos can explain the lightness of SM neutrinos. Richer dark-sector models can generate the baryon-antibaryon asymmetry of the universe and potentially address the hierarchy problem ({\em e.g.}, in relaxion models). Dark sectors can address the strong-CP problem via axion or axion-like particles~(ALPs). Finally, dark sectors can also ameliorate several anomalies in data, such as the anomalous magnetic moment of the muon, $(g-2)_\mu$, and short-baseline neutrino anomalies.  These many directions, as well as DM, can be realized by \textbf{BSM physics below the electroweak~(EW) scale}, which can be systematically classified and explored based on the {\em portals} mediating interactions between the SM and a low-mass dark sector.  Because the new physics of dark sectors is at low mass and weakly coupled, the most powerful means to explore it is through intensity-frontier experiments, with beam intensity, precise instrumentation, and detector geometry as paramount sensitivity drivers, rather than beam energy.

\textbf{Maximizing the possibility of discovering a dark sector requires a four-pronged approach.} Support for dark-sector analyses at multi-purpose experiments, the DMNI program, an expansion of DMNI to include a focus on complementary signatures such as with visible final states (especially long-lived particles), and dark-sector theory will enable a powerful and broad exploration of the dark sector in the coming decade, including comprehensive coverage of major milestones in thermal DM, dramatic increases in sensitivity to vector and (pseudo)scalar mediators decaying (semi)visibly, as well as the full exploration of models addressing some anomalies in data.   
Large multi-purpose detectors, especially Belle-II and LHCb, can efficiently cover large regions of parameter space for both GeV-scale DM and mediators to a dark sector that decay to ordinary matter. These important sensitivity gains are inexpensive, requiring only a few physicists to work on trigger optimization and analysis.  
Other important regions of parameter space are best accessed with smaller specialized experiments.  Through the competitive DMNI program, the community and DOE selected 
two intensity-frontier projects, CCM200 and LDMX, to explore low-mass thermal DM milestones with different timescales and complementary sensitivity.
The CCM200 proton-beam re-scattering experiment was completed and commissioned in 2021 and is now operating, whereas   
the LDMX electron-beam missing-momentum experiment has thus far only received pre-project funds and awaits construction funding. 
LDMX offers a unique and powerful opportunity at low cost to explore low-mass DM, including fully exploring the low-mass thermal DM region, which makes timely completion of the DMNI program crucial. 
While multi-purpose and DMNI experiments enable great discovery potential, many well-motivated scenarios will not be covered, including long-lived dark-sector particles decaying to visible SM
particles. 
Therefore, the US community should also seize the opportunity to lead new dedicated high-intensity experiments searching for visible decays of long-lived dark-sector particles (\textit{Thrust 2} in the DMNI report), by selecting among several proposals that would greatly expand sensitivity in this arena. This can be achieved in future rounds of DMNI.
In addition, it will be important to further expand the experimental program to provide complementary sensitivity, {\em e.g.}, to scenarios where dark-sector mediators have flavor-specific couplings. 
To summarize: targeted investment 
in specialized experiments will enable rapid progress and provide tremendous discovery potential.
Finally, continued support for leadership in dark-sector theory will also be critical both to continue developing dark-sector models to address open problems in particle physics and cosmology, and to maximize the efficacy of the experimental program, where the track record of theorists pioneering new approaches is strong. 

\textbf{Objectives and structure of this report.} This report summarizes the scientific importance of and motivations for searches for dark-sector particles below the EW scale, the current status and recent progress made in these endeavors, the landscape and major milestones motivating future exploration, and the most promising and exciting opportunities to reach these milestones over the next decade. 
We summarize the different experimental approaches 
and we discuss proposed experiments and their accelerator facilities.
In addition, as part of the Snowmass process, we defined three primary research areas, each with associated ambitious---but achievable---goals for the next decade. This categorization is motivated, in part, by how we search for DM in different scenarios. 
When DM is light, portals to the dark sector allow its production and detection at accelerators
({\em e.g.}, in mediator decay if the DM is lighter than half of the mediator mass, or coupled through an off-shell mediator). In fact, accelerators can probe DM interaction strengths motivated by thermal freeze-out explanations for  the cosmological abundance of DM. 
If DM is heavier,  the mediator decays into visible SM particles. In addition to thermal DM models, visible mediators also arise in theories that address various open problems in particle physics ({\em e.g.}, the strong-CP problem, neutrino masses, and the hierarchy problem). 
A third scenario is where the dark sector is richer, which can lead to decays of the mediator to both DM and SM particles, or to other final states not considered in the standard minimal benchmark models.
Each of these research areas is discussed in detail in this report.

\begin{figure}[b!]
    \centering
    \includegraphics[width=0.7\textwidth]{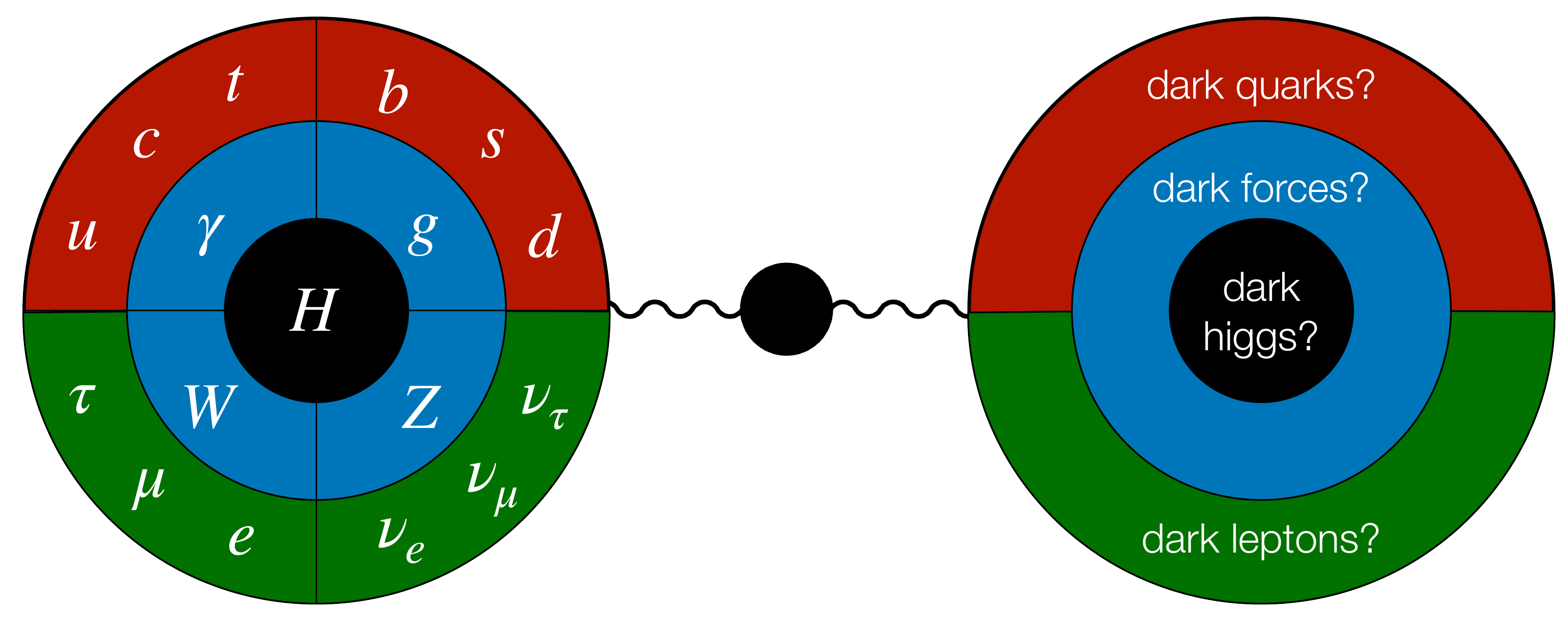}
    \caption{Cartoon schematic of the dark-sector paradigm. The same complexity observed in ordinary matter, as described by the Standard Model, may also be present in the dark sector. 
    Interactions between the Standard Model and the dark sector can arise via the so-called portal interactions.} 
    \label{rf6-fig:dark-sector-cartoon}
\end{figure}

\begin{center}
\textbf{Theoretical Framework} 
\end{center}
\vspace{-1em}
The leading possible interactions between ordinary and dark-sector particles, classified below, are known as portals. The strength of portal interactions can be naturally suppressed by symmetry reasons, and can arise only at higher orders in perturbation theory. 
Figure~\ref{rf6-fig:dark-sector-cartoon} shows a schematic representation of the dark-sector paradigm.
This simple scenario where dark-sector particles only couple indirectly to ordinary matter naturally leads to feeble interactions, and opens the door to the possibility that BSM physics may exist below the EW scale. In fact, the mass of dark-sector particles might be naturally light if protected by some symmetry (this is the case, {\em e.g.}, for ALPs).
In addition, the inherently feeble interactions of dark-sector matter with ordinary matter provides a natural thermal-production origin for DM for the case where DM is light, extending the well-known {\em WIMP miracle} to lower mass scales. Due to the Lee-Weinberg bound, light mediators are generically needed if DM is at or below the GeV scale.
Therefore, testing the dark-sector hypothesis requires innovative high-intensity experiments, not necessarily high energies.

The landscape of potentially viable dark sectors is broad with many regions largely untested experimentally and unexplored theoretically. 
Even so, the physics of dark sectors can be systematically studied using the few allowed portal interactions as a guide. The gauge and Lorentz symmetries of the SM greatly restrict how the dark-sector mediators can couple to ordinary matter.
The dominant interactions between SM particles and dark-sector mediators are naturally those with the lowest dimensions:

\textbf{Vector Portal}: 
DM particles may interact with other DM particles via a dark force similar to the electromagnetic force felt by ordinary matter.
The {\em dark photon}, $A'$, that mediates this force can obtain a small coupling to the electromagnetic current due to kinetic mixing between the SM hypercharge and $A'$ field strength tensors via the operator $(\varepsilon/2)F^{\mu\nu} F'_{\mu\nu}$, where $\varepsilon$ characterizes the size of the kinetic mixing. 

\textbf{Higgs Portal}: 
The {\em dark scalar}, $S$, coupled to the gauge-invariant Higgs mass operator, $(\mu S + \lambda S^2) H^\dag H$, can mix with the SM Higgs boson, with a mixing angle denoted by $\theta$. 
The $S$ boson then inherits the interactions of the Higgs to SM particles suppressed by this mass mixing.

\textbf{Neutrino Portal}:
A gauge-singlet fermion, $N$, called a {\em heavy neutral lepton} (HNL), can couple to the gauge-invariant operator formed of the SM lepton and Higgs $SU(2)$ doublets, $yNH L$. 
Following EW symmetry breaking, the HNLs mix with the SM neutrinos. 
HNLs can be Majorana, Dirac, or pseudo-Dirac particles.

\textbf{Axion Portal}: 
ALPs, $a$, are pseudoscalar particles whose couplings to the SM gauge bosons are highly suppressed at low energies by a large decay constant, $f_a$, via operators like $aF_{\mu\nu}\tilde F^{\mu\nu}/f_a$.
ALPs are pseudo-Nambu-Goldstone bosons; therefore, their masses, $m_a$, are expected to be $m_a \ll f_a$.

The first three portals are renormalizable, while the axion portal is a dimension-5 operator suppressed by the high-energy scale $f_a$. 
Direct coupling of new gauge bosons to SM fermion vector currents are also possible, though most are anomalous.
The few that are not anomalous, {\em e.g.}\ $B\!-\!L$, $L_{\mu}\!-\! L_{\tau}$, and $B\!- \!3L_{\tau}$, are viable and interesting ways to connect ordinary matter to a dark sector. 
This completes the list of possible dimension-5 and below interactions between a dark sector and the SM. 
This small list of possible interactions sharpens the focus of the experimental discovery effort. Furthermore, as we will highlight in this report, targets for the size of the couplings between the dark sector and the SM arise from phenomenological and/or theoretical motivations. We aim to reach these targets with the experimental program described in this report.

\begin{center}
\textbf{Experimental Landscape} 
\end{center}
\vspace{-1em}
A large experimental and theoretical community effort has led to much progress over the last decade in the search for MeV-to-GeV mass dark-sector particles. 
The first results were obtained from re-analyses of data sets of past experiments, including many that were not designed to search for dark sectors.
Next, searches were performed using large multi-purpose experiments, 
where again the first results were obtained using data collected 
with other purposes in mind. 
A number of specialized smaller-scale experiments began running during the past decade, with some having published results. 
Most recently, two intensity-frontier experiments (LDMX and CCM200) received some support 
from the DOE-supported DMNI program.

Moving forward, rapid progress towards achieving the goals discussed below is possible if the US commits to fully exploiting large multi-purpose detectors, and to targeted investment in specialized experiments. 
A diverse program of experiments, aimed at achieving these ambitious goals, has tremendous discovery potential. Even null results will have a huge impact by dramatically narrowing the range of viable DM candidates and well-motivated dark-sector scenarios.
The experimental landscape is discussed in detail in Sec.~\ref{rf6-sec:experiments} of this report and in a dedicated Snowmass white paper \cite{Ilten:2022lfq}.

\begin{center}
\textbf{Dark matter production at intensity-frontier experiments}
\end{center}
\vspace{-1em}
As discussed above, matter from a dark sector is an attractive DM candidate.  Indeed, thermal freeze-out of light DM (below the proton mass) requires DM interactions through a light force carrier with feeble SM couplings, which arise naturally in a dark sector.  Taken together, these arguments motivate an exciting program to search for light DM by producing it at intensity-frontier experiments.  

\begin{figure}[p!]
\begin{center}
\includegraphics[width=0.9\textwidth]{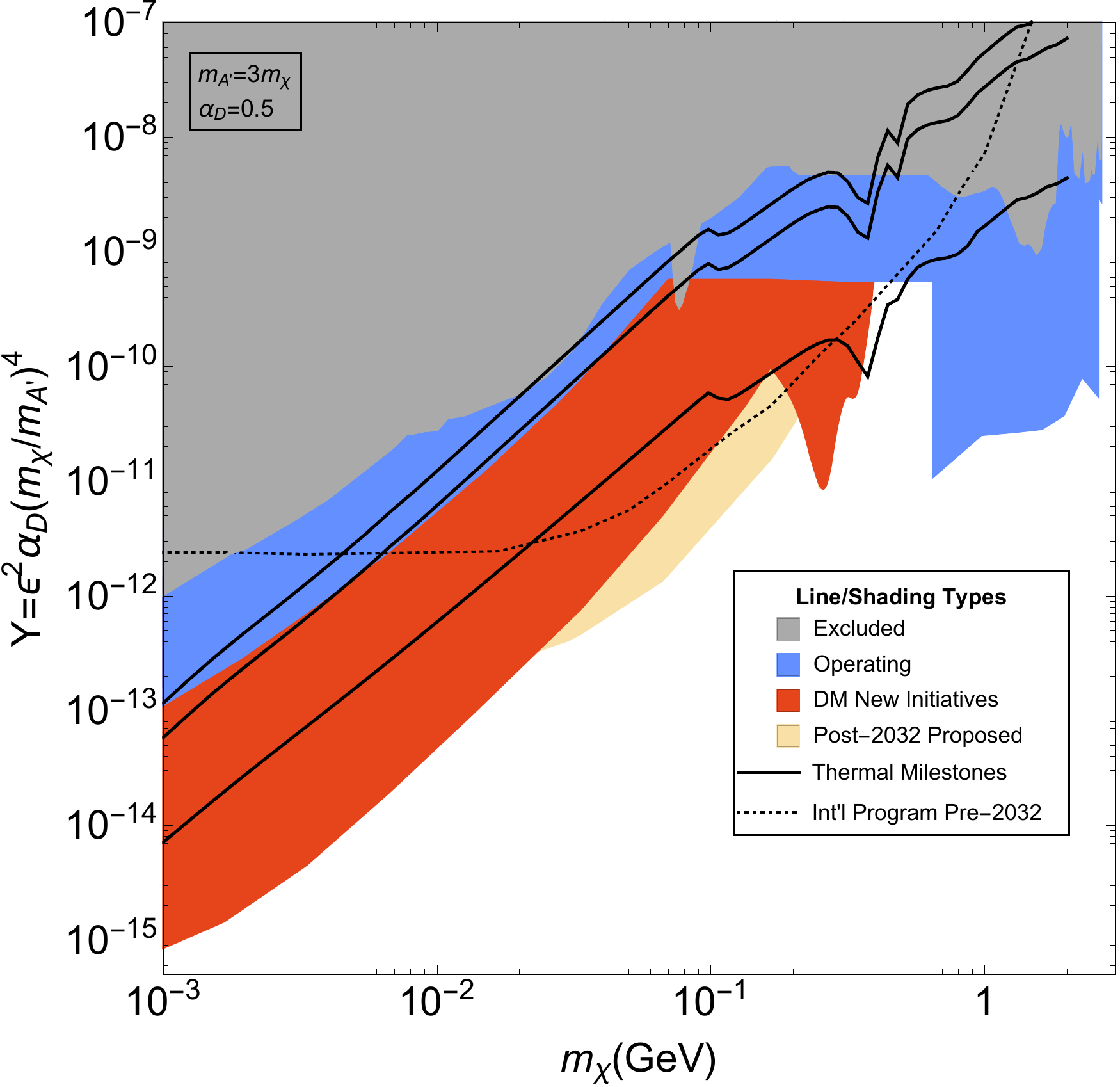}
\end{center}
\caption{
Dark matter production at accelerators~\cite{Krnjaic:2022ozp}: 
Thermal milestones for the kinetically mixed dark photon model, a key light DM benchmark (shown as black solid lines), along with exclusions from past experiments (gray regions), projected sensitivities of future projects that are operating or have
secured full funding (blue regions), and the experiments partially funded by the Dark Matter New Initiatives~(DMNI) program (red regions).
Primarily international experiments are shown as a dashed line.
Proposed experiments that are farther into the future are shown in light yellow.  
As can be seen, the combination of operating and DMNI experiments can comprehensively explore the thermal DM targets, considered a major milestone of dark-sector physics. 
See Fig.~\ref{rf6-fig:phys-BI1} for a more detailed version of this figure showing individual experiments color-coded according to which SM fermion coupling is employed to produce the DM. In variations of this model, some species may have suppressed couplings, making different SM fermion couplings complementary in this extended model space.
}
\label{rf6-fig:phys-BI1-cartoon}
\end{figure}

A particularly exciting opportunity is presented by the vector portal, where the simplest models of thermal freeze-out relate the cosmological abundance of DM to the signals expected at accelerator-based experiments, defining a sharp and high-priority milestone in DM interaction strength as a function of its mass.  This milestone (illustrated by the black diagonal lines in Fig.~\ref{rf6-fig:phys-BI1-cartoon}) is not yet experimentally constrained over most of the MeV-to-GeV mass range. However, at interaction strengths 10 to 1000 times smaller than those presently explored, it is within reach of next-generation experiments.

Accelerator-based production of DM particles can be observed at intensity-frontier experiments, including dedicated fixed-target experiments, downstream detectors, and flavor factories.  
Three categories of search strategy, with complementary sensitivity, are employed:
(i) inferring missing energy, momentum, or mass;
(ii) detecting re-scattering of DM particles in downstream detectors;
(iii) observing semi-visible signatures of metastable dark-sector particles. 

In the next decade, the primary goal will be to explore parameter space motivated by thermal DM models, the dark-sector paradigm, and anomalies in data.
Figure~\ref{rf6-fig:phys-BI1-cartoon} shows that the milestone highlighted above---full exploration of the range of interaction strengths compatible with light DM thermal freeze-out via the simplest mechanism of $s$-channel annihilation to SM particles mediated by a dark photon (black diagonal lines in Fig.~\ref{rf6-fig:phys-BI1-cartoon})---is achievable by near-future experiments. 
While for this one goal many future experiments seem redundant (see Fig.~\ref{rf6-fig:phys-BI1}), 
the use of multiple complementary techniques is important:
(i) to probe a broader class of thermal freeze-out models, such as those where a mediator, unlike the dark photon, does not couple to electrons; (ii) to test
models where meta-stable particles in the dark sector play important roles in DM cosmology and enable new discovery techniques;
(iii) to explore neutrino portal annihilation, which has qualitatively different experimental signals. 
These opportunities are highlighted and examined in more detail in Sec.~\ref{rf6-sec:BI1} of this report and in a dedicated Snowmass white paper~\cite{Krnjaic:2022ozp}. 

In all of these scenarios, DM production at accelerators offers a window to DM candidates that are impossible to discover in other ways.  Accelerator-based light-DM searches target lower DM masses than conventional direct detection experiments, and explore models where mediators decaying to SM particles (see next research area) are typically absent. Accelerators can also explore scenarios in which DM is produced with an excited metastable state. These searches are also complementary to the low-threshold direct-detection experiments, which probe sub-GeV DM at vastly different kinematics: direct detection probes non-relativistic scattering, whereas accelerators explore semi-relativistic DM production.  This translates into complementary discovery potential. Accelerators are optimal for discovering DM whose interactions are suppressed at low velocities, including thermal freeze-out through a dark photon with generic spin and mass structure. In contrast, low-threshold direct detection is a powerful probe for interactions that are enhanced at low velocities, including models of freeze-in through a light mediator.

\begin{center}
\textbf{Exploring dark-sector portals with high-intensity experiments}
\end{center}
\vspace{-1em}
The minimal portals provide a variety of well-motivated DM scenarios with novel cosmology and phenomenology. 
For example, the {\em secluded DM} scenario, where heavy DM is produced thermally in the early universe via its annihilation to lighter mediator particles, can be viable in any of the minimal portals. 
The requirement of thermalization of the SM and dark sector imposes a lower bound on the strength of the portal coupling, thus defining a target region for high-intensity experiments. 
The minimal portals may also play an important role in solutions to puzzles motivated, {\em e.g.}, by naturalness considerations or the strong-CP problem. In these scenarios, the mediator decays to visible SM particles.
For more detailed discussion on the connections between the minimal portal models and the potential answers to the big questions in particle physics and cosmology, see Sec.~\ref{rf6-sec:BI2} of this report and the dedicated Snowmass white paper \cite{Batell:2022dpx}.

The focus here is on minimal extensions of the SM featuring a single new light mediator coupled through one of the portal interactions. 
In this case, the portal coupling governs both production and decay of the mediator.  Beyond minimal models, the same experimental signals can arise if DM (or other dark-sector matter) is heavier than half the mediator mass. For these reasons, searching for visible decays of dark-sector mediators should be a high-priority goal of the dark-sector high-intensity program.
Given that dark sectors in this category produce visible signatures, they can be pursued on the following experimental fronts: 
electron and proton beam fixed-target experiments provide excellent reach at low masses over a broad range of couplings;  
medium-energy $e^+ e^-$ colliders are sensitive to moderate couplings both at low and intermediate masses; 
precision studies of meson and lepton decays, including those at 
muon facilities, probe low masses and small couplings;
and existing experiments and proposed long-lived-particle (LLP) detectors at the LHC make accessible regions of parameter space at low masses and relatively small couplings.

\begin{figure}[p!]
\begin{center}
\includegraphics[width=0.9\textwidth]{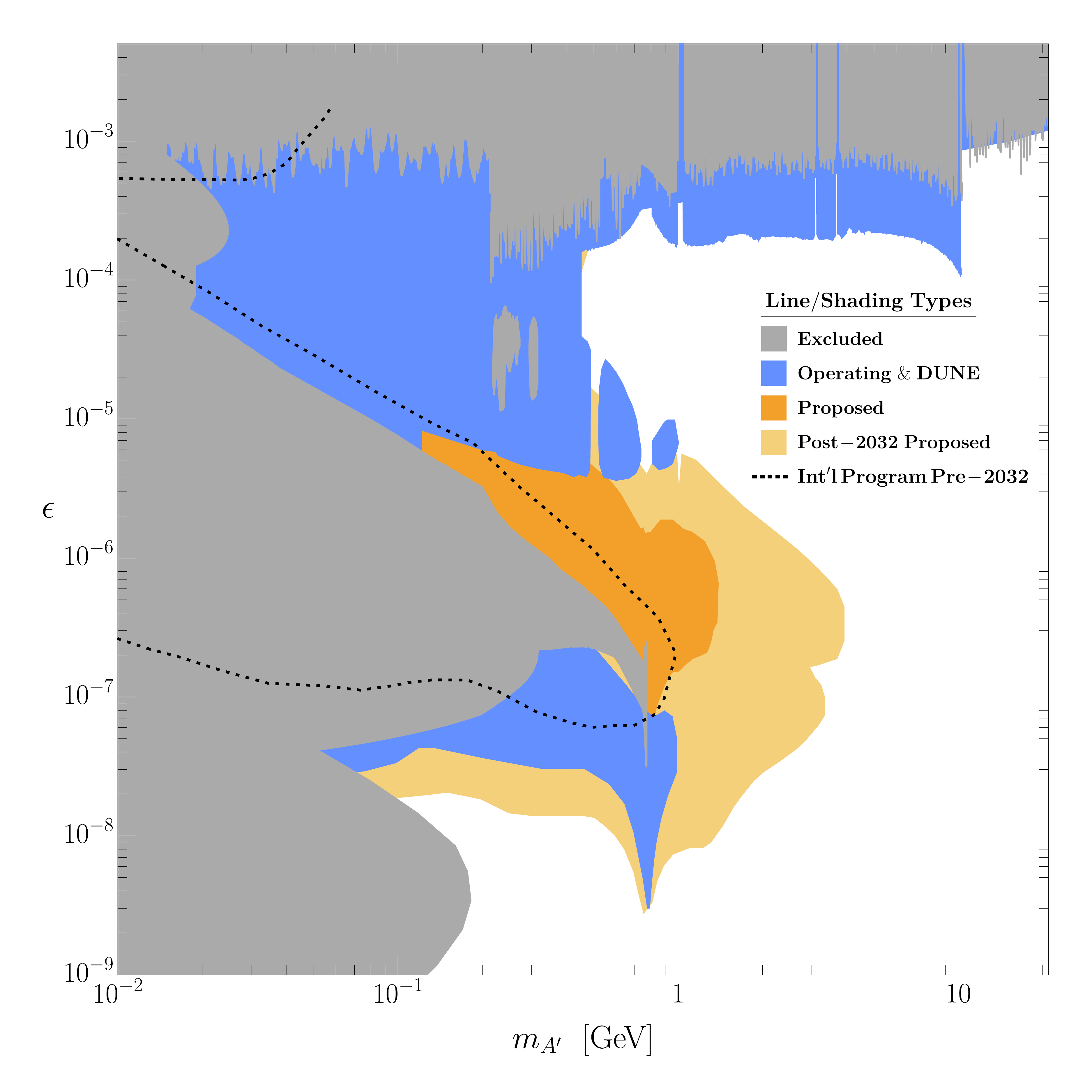}
\end{center}
\caption{ 
Visible dark-photon parameter space (compatible with secluded thermal DM)~\cite{Batell:2022dpx}: 
Near-term and future opportunities to search for visibly decaying dark photons interacting through the vector portal displayed in the dark photon mass $(m_{A'})$ -- kinetic mixing  $(\epsilon)$ parameter space. 
Constraints from past experiments (gray  regions) and projected sensitivities from operating and fully funded experiments and DUNE (blue regions),  and other proposed near-term (pre-2032) experiments based in the US and/or with strong US leadership (orange region) are shown. 
Primarily international projects are shown as a dashed line.
Proposed experiments that are farther into the future are shown in light yellow. 
See Fig.~\ref{rf6-fig:phys-BI2-aprime} for a more detailed version of this figure showing individual experiments color-coded by experimental approach, highlighting one aspect of the complementarity between different facilities/experiments. Collectively, these experiments are poised to cover large regions of open dark-photon parameter space. 
}
\label{rf6-fig:phys-BI2-aprime-cartoon}
\end{figure}

\begin{figure}[p!]
\begin{center}
\includegraphics[width=0.9\textwidth]{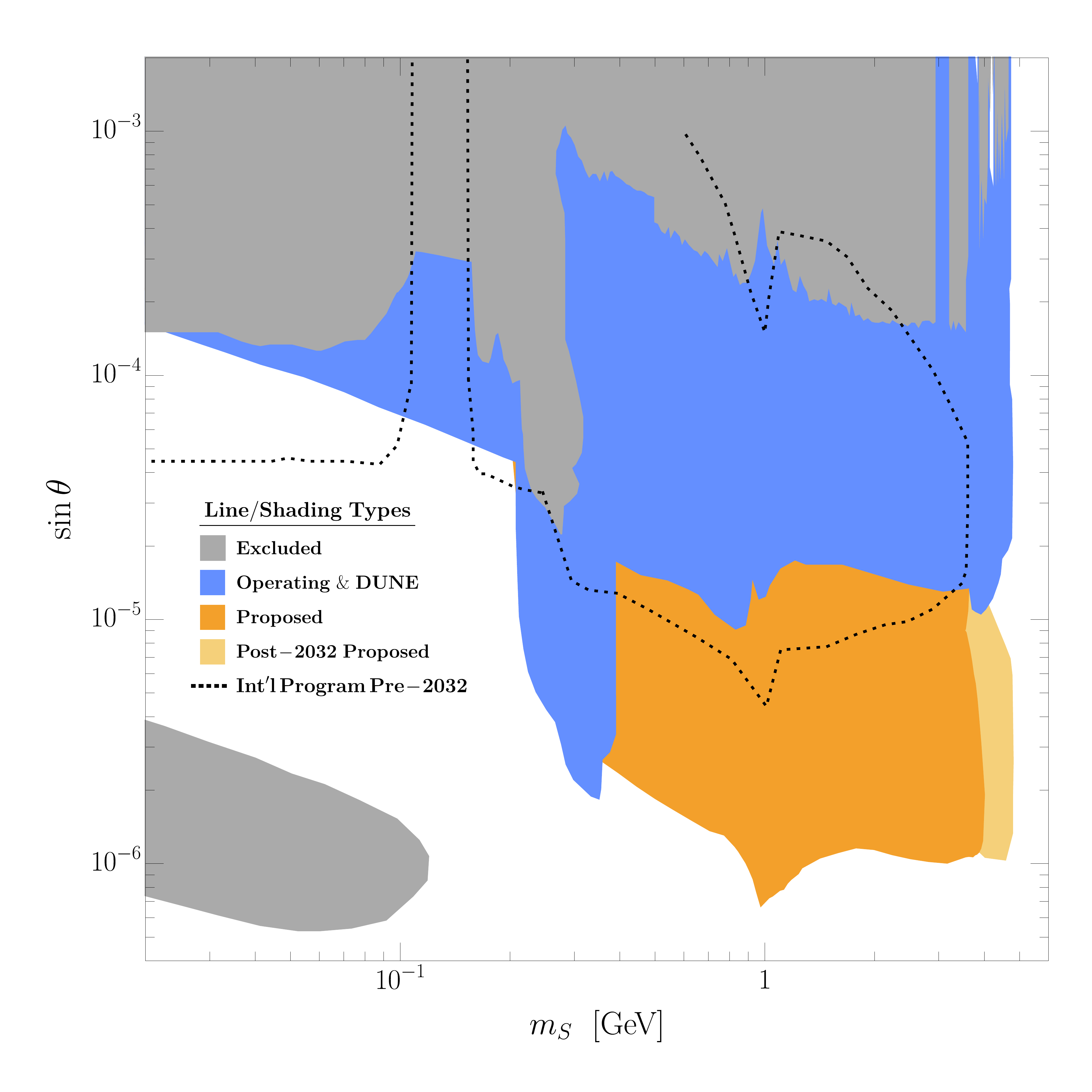}
\end{center}
\caption{
Visible dark-scalar parameter space (compatible with secluded thermal DM) ~\cite{Batell:2022dpx}: 
Near-term and future opportunities to search for visibly decaying dark scalars interacting through the Higgs portal displayed in the scalar mass $(m_{S})$ -- mixing angle $(\sin\theta)$ parameter space. 
Constraints from past experiments (gray  regions) and projected sensitivities from operating and fully funded experiments and DUNE (blue regions), and other proposed near-term (pre-2032) experiments based in the US and/or with strong US leadership (orange region) are shown. 
Primarily international projects are shown as a dashed line.
Proposed experiments that are farther into the future are shown in light yellow.
See Fig.~\ref{rf6-fig:phys-BI2-higgs} for a more detailed version of this figure showing individual experiments color-coded by experimental approach, highlighting one aspect of the complementarity between different facilities/experiments. 
Collectively, these experiments are poised to cover large regions of open dark-scalar parameter space.
}
\label{rf6-fig:phys-BI2-higgs-cartoon}
\end{figure}

Reference~\cite{Batell:2022dpx} demonstrates the exciting opportunity to explore vast uncharted territory of all four portal interactions discussed above
during the next decade and beyond. 
Figure~\ref{rf6-fig:phys-BI2-aprime-cartoon} shows the current constraints and future projected sensitivities for the minimal dark-photon model.
The parameter space shown in Fig.~\ref{rf6-fig:phys-BI2-aprime-cartoon} is all well-motivated since it leads to thermal production of secluded DM. Therefore, it is desirable to explore as much of this space as possible. 
Some regions of this parameter space also have additional motivation, {\em e.g.}, if the kinetic mixing is induced at the one or two-loop level, $\varepsilon$ is expected to be in the $10^{-6}$ to $10^{-3}$ range and potential solutions to the core/cusp problem motivate an MeV-to-GeV mediator mass. 
As can be seen, operating experiments will explore all of the open parameter space in the region $m_{A'} \lesssim 0.5$\,GeV and $\epsilon \gtrsim 10^{-5}$. A next-generation proton fixed-target beam dump experiment at Fermilab (DarkQuest) could explore $\epsilon$ values more than an order of magnitude smaller and masses up to the GeV scale, while a $pp$ forward LLP experiment (FASER2) that utilizes high-luminosity LHC data can probe a similar parameter space (the combined region explored by these two experiments is shown in orange in Fig.~\ref{rf6-fig:phys-BI2-aprime-cartoon}). 
Figure~\ref{rf6-fig:phys-BI2-higgs-cartoon} shows the corresponding parameter space for the minimal dark-scalar model.
We see that operating experiments will greatly extend our current capabilities, but also that future LLP experiments at CERN (orange region in Fig.~\ref{rf6-fig:phys-BI2-higgs-cartoon})
and a next-generation kaon experiment 
are needed to explore smaller couplings (see Fig.~\ref{rf6-fig:phys-BI2-higgs} for more details on the regions covered by specific experiments). 
Additional motivations and other portals are discussed briefly in Sec.~\ref{rf6-sec:BI2} of this report and in much more detail in \cite{Batell:2022dpx}.

\begin{center}
\textbf{New flavors and rich structures of the dark sector at high-intensity experiments}
\end{center}
\vspace{-1em}
Experimental work on dark sectors has been primarily focused on minimal scenarios, typically with a single mediator, a single DM candidate, and the assumption of flavor universality. 
However, dark sectors could have non-minimal structure, similar in complexity to the SM,  which can lead to a far richer phenomenology---and may require new experimental strategies. 
Many theoretically and/or experimentally well-motivated dark-sector models lead to interesting non-minimal dark-sector phenomena that can be efficiently searched for in high-intensity experiments. A set of novel signatures are motivated by either
data-driven anomalies ({\em e.g.} $(g-2)_\mu$), 
theoretical problems ({\em e.g.}\ the nature of DM, the flavor puzzle), 
or by the desire to achieve more complete coverage of the standard benchmark models beyond the assumption of minimality. 

\begin{figure}[p!]
    \centering
     \includegraphics[width=0.9\textwidth]{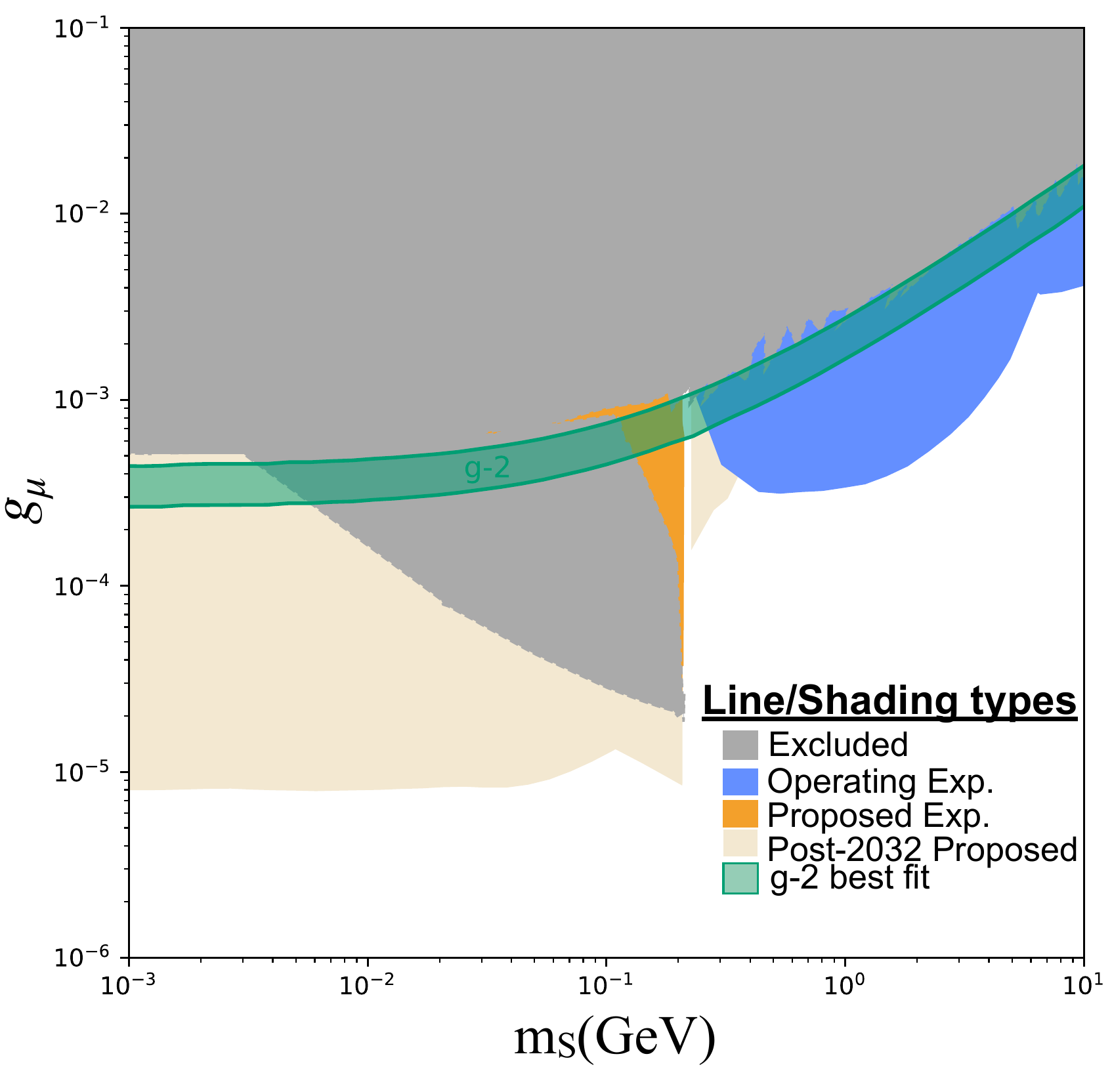}
    \caption{Muon-philic scalar model (one of the few remaining possible dark-sector explanations of the $(g-2)_\mu$ anomaly)~\cite{Harris:2022vnx}: The parameter space that can explain the $(g-2)_\mu$ anomaly is shown as a green band and represents the main experimental target of this model.
    Constraints from past experiments (gray  regions) and projected sensitivities from operating and fully funded experiments (blue regions), other proposed near-term (pre-2032) experiments based in the US and/or with strong US leadership (orange region),
and proposed future (post-2032) experiments (light yellow regions) are shown. The combination of running/fully funded experiments and proposed near-term experiments will nearly be able to fully cover the parameter space of muon-philic scalars with a mass above a few MeV able to address the anomaly.
    See Fig.~\ref{rf6-fig:phys-BI3-g-2} for a more detailed version of this figure showing individual experiments color-coded by experimental approach. 
   } 
   \label{rf6-fig:phys-BI3-g-2-cartoon}
 \end{figure}

Dark-sector models can address 
a range of data-driven anomalies. Several examples are discussed in detail in a dedicated Snowmass white paper~\cite{Harris:2022vnx}. 
In the case of the $(g-2)_\mu$ anomaly, in the last decade, high-intensity experiments were able to completely rule out the simplest dark-sector explanations. The goal for the coming years will be to completely probe non-minimal dark-sector models able to address this anomaly. As discussed in \cite{Harris:2022vnx}, only a few models are still unexplored. These models
involve non-minimal flavor-dependent interactions, and can largely be explored in the next decade by high-intensity experiments, either ruling out dark-sector explanations or making an historic discovery. An example scenario is shown in Fig.~\ref{rf6-fig:phys-BI3-g-2-cartoon}, where the $(g-2)_\mu$-favored region of parameter space will only be partially probed with operating and fully funded experiments. To achieve complete coverage, new proton fixed-target experiments  will be needed.

As an example of theory-based motivations, the dark-sector framework can include generalizations of the QCD axion referred to as ALPs. 
The high-intensity dark-sector program is able to probe all ALP couplings to SM particles for a broad range of masses. 
Reference~\cite{Harris:2022vnx} highlights the case of a flavor-violating QCD axion model, where the parameter space in which the QCD axion is a viable cold DM candidate can be explored. 

\begin{figure}[p!]
\centering
\includegraphics[width=0.9\textwidth]{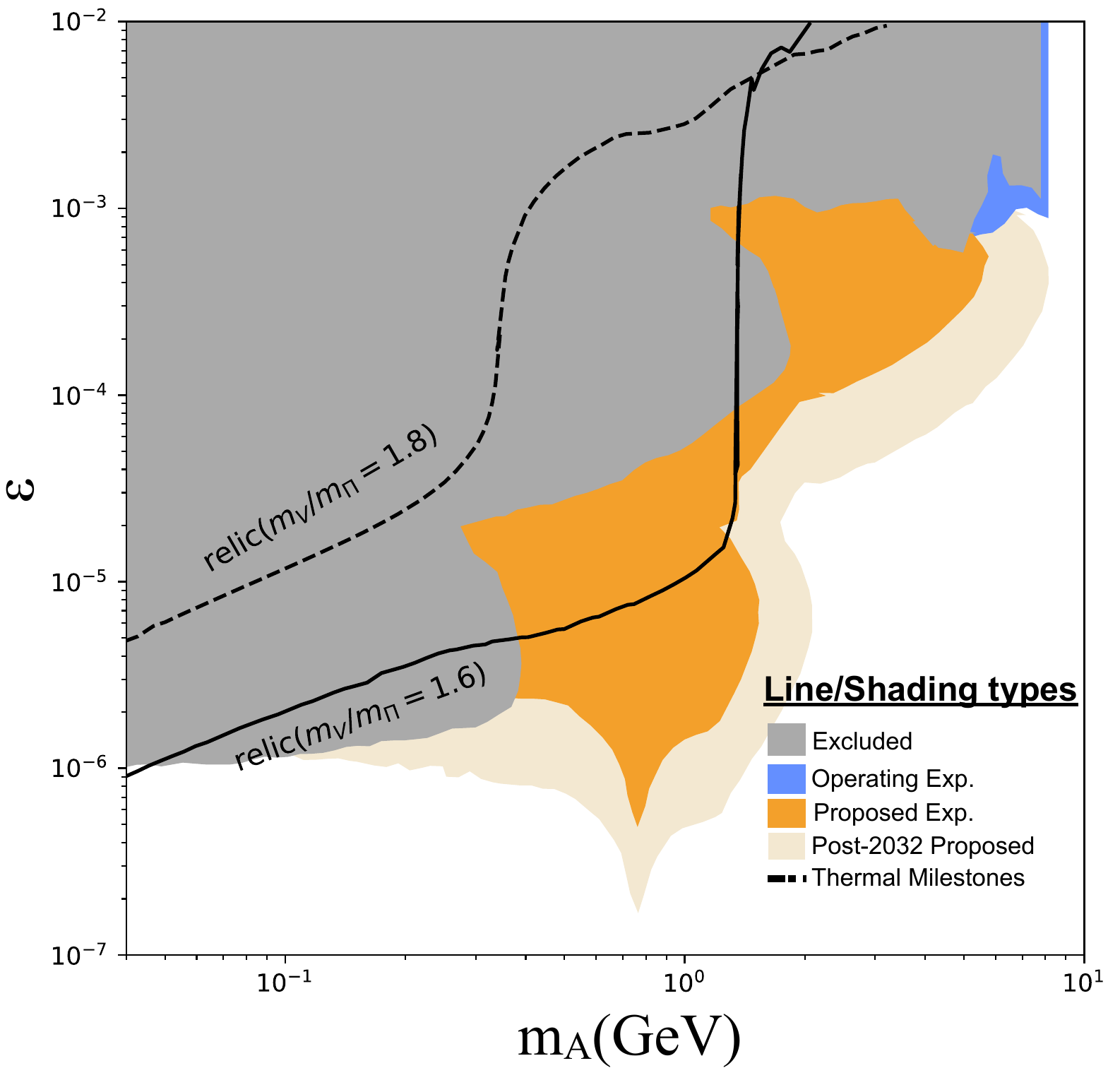}
\caption{An example of the SIMP parameter space for a QCD-like strongly coupled dark sector gauged under a new $U(1)$ with a dark photon with interactions to the SM through the mixing parameter $\epsilon$ (see \cite{Harris:2022vnx} for details about the dark-sector model parameters).
Constraints from past experiments (gray  regions) and projected sensitivities from operating and fully funded experiments (blue regions), other proposed near-term (pre-2032) experiments based in the US and/or with strong US leadership (orange region),
and proposed future (post-2032) experiments (light yellow region) are shown.
 Dark pions constitute all of the observed DM abundance on solid (dashed) black
  contours, while DM is overabundant below these lines. Near-term proposed experiments will be able to fully cover these thermal targets.
  See Fig.~\ref{rf6-fig:phys-BI3-simp} for a more detailed version of this figure showing individual experiments color-coded by experimental approach. 
\label{rf6-fig:phys-BI3-simp-cartoon}}
\end{figure}

Finally, thermal DM models where the DM belongs to a dark sector often involve the presence of excited DM states. 
Reference~\cite{Harris:2022vnx} categorizes these extended models according to whether the dark-sector mass scale arises in the weak- or strong-coupling regime. 
Examples from each category, namely inelastic dark matter (iDM) and strongly interacting massive particles (SIMPs), are discussed in detail, with both predicting a wide range of new phenomena with semi-visible displaced decays of the DM excited states. Interestingly, the mechanism to deplete the DM abundance in SIMP models hints at a DM mass below the GeV scale.
In both models, the requirement of reproducing the observed DM relic density defines experimental targets. High-intensity experiments
in the next decade will be able to probe most of this parameter space. 
Figure~\ref{rf6-fig:phys-BI3-simp-cartoon} shows this for the SIMP model (see the black solid and dashed lines for the thermal targets). The thermal-DM region not covered by previous and operating experiments can be explored by a future proton fixed-target experiment (DarkQuest). 
More details can be found in Sec.~\ref{rf6-sec:BI3} of this report and in \cite{Harris:2022vnx}.

\begin{figure}[b!]
    \centering
    \includegraphics[width=0.99\textwidth]{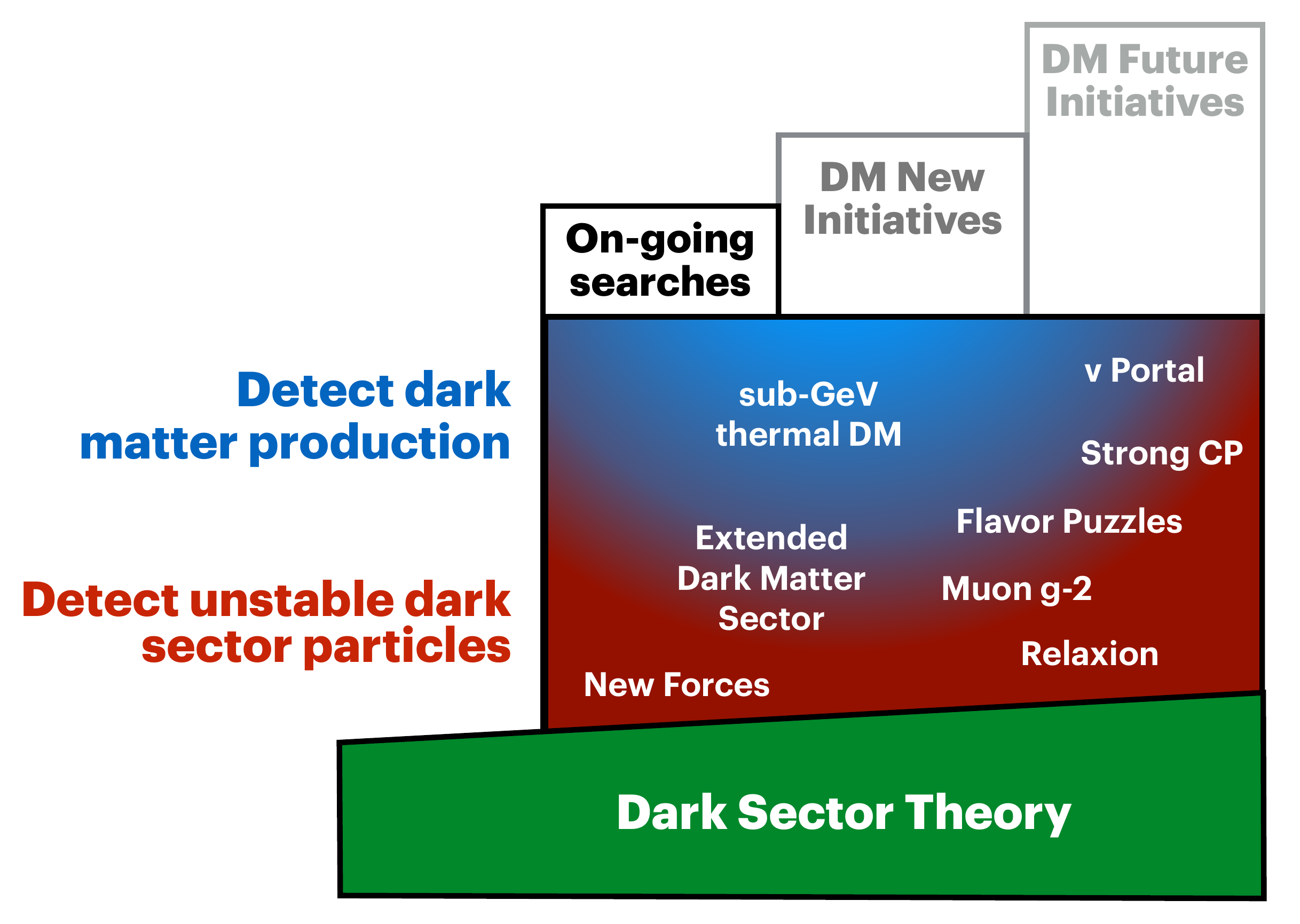}
    \caption{Schematic representation of the RF6 roadmap.}
    \label{rf6-fig:flag}
\end{figure}

\begin{center}
\textbf{Scientific Opportunities \& Roadmap}
\end{center}
\vspace{-1em}
The goals presented above are complementary and must be pursued in parallel to fully explore the physics of dark sectors. 
To realize these goals over the next decade, the US high-intensity dark-sector community must do the following:
\begin{itemize}[leftmargin=1.0em]
    \item \textbf{Exploit the capabilities of existing large multi-purpose detectors},  especially Belle-II and LHCb. These experiments cover large regions of the most compelling target parameter space in Figs.~\ref{rf6-fig:phys-BI1}--\ref{rf6-fig:phys-BI3-g-2}, and also for many of the other well-motivated scenarios highlighted in Secs.~\ref{rf6-sec:BI1}--\ref{rf6-sec:BI3} of this report. 
    The additional investment required to do world-leading dark-sector physics at these experiments is small, typically just a few physicists to write the trigger/real-time selections and to analyze the data. 
    \item \textbf{Invest in the completion of the DMNI program}, in particular fully funding the LDMX experiment which received pre-project funds through DMNI and is awaiting construction funding. 
    As can be seen in Fig.~\ref{rf6-fig:phys-BI1}, LDMX provides a unique opportunity to fully explore the low-mass thermal-DM region (at low cost).  
    The community and the DOE have already chosen to pursue the LDMX experiment, and it will be important to follow through on this initial investment (in a timely manner). 
    \item The first two items will enable great discovery potential. However, many well-motivated scenarios will not be covered. Crucially, this includes models that predict dark-sector particles decaying to visible SM particles, especially in the long-lived regime 
    (see Figs. \ref{rf6-fig:phys-BI2-aprime}--\ref{rf6-fig:phys-BI3-simp}). 
    The goal of exploring this parameter space,  a vital component of a robust dark-sectors program, was presented as \textit{Thrust 2} in the DMNI report, though no support was allocated thus far towards this goal through DMNI. 
    Therefore, \textbf{DMNI funding must be expanded to achieve the goals laid out in the DMNI report and to reach the full potential of the accelerator-based dark-sector program}. 
    In particular, a number of experiments have been proposed to explore the (semi)visibly decaying long-lived regime, both at US-based accelerators and at the LHC. 
    The US dark-sectors community should select which of these exciting ideas to fund in a second DMNI round. 
    In addition, in future rounds of DMNI it will be important to further expand the experimental program to provide complementary sensitivity, {\em e.g.}, to scenarios where dark-sector mediators have flavor-specific couplings. 
    To summarize: targeted investment in specialized experiments will enable rapid progress and provide tremendous discovery potential. The US community must select a portfolio of well-motivated, unique, affordable, and mutually complementary intensity-frontier dark-sector experiments to lead during the next decade (a range of exciting opportunities are presented in this report). 
    
    \item \textbf{Support dark-sector theory} efforts to: better understand which dark-sector scenarios can address (current and future) open problems in particle physics; develop new ideas for exploring the dark sector; and collaborate at every stage of new dark-sector experiments, from design through interpretation of the data. This type of theory work has been at the foundation of essentially all ongoing and planned experimental activities in this growing field. Support for theory-experiment collaboration and workshops, analogous to the aims of the FIP Physics Center at CERN, will be important. Such support (focused on US efforts) had been envisioned through the DMNI program but was cancelled due to the pandemic in 2020. Restarting this is appropriate.     
\end{itemize}
Collectively, these activities provide a roadmap (illustrated in Fig.~\ref{rf6-fig:flag}) for the US high-intensity dark-sectors program for the next decade.
Discovering a dark sector would not only shed light on the enigma of dark matter and several additional open issues in particle physics---but would revolutionize our understanding of the universe and open an entire new pathway to studying Nature. 
Following the roadmap outlined here will secure the leadership role of the US dark-sectors community.

\clearpage

\section{Experiments and facilities} 
\label{rf6-sec:experiments}

\textit{This section is adapted from the executive summary of a Snowmass white paper solicited by the RF6 conveners and edited by Phil Ilten and Nhan Tran~\cite{Ilten:2022lfq}.} 

\input{experiments}

\clearpage 

\section{Dark matter production at intensity-frontier experiments}
\label{rf6-sec:BI1}

\textit{This section is adapted from the executive summary of a Snowmass white paper solicited by the RF6 conveners and edited by Gordan Krnjaic and Natalia Toro~\cite{Krnjaic:2022ozp}.}

\input{big-idea1}

\section{Exploring Dark Sector Portals with High Intensity Experiments}
\label{rf6-sec:BI2}

\textit{This section is adapted from the executive summary of a Snowmass white paper solicited by the RF6 conveners and edited by Brian Batell, Nikita Blinov, Christopher Hearty, and Robert McGehee~\cite{Batell:2022dpx}.}

\input{big-idea2}

\clearpage

\section{New flavors and rich structures in dark sectors}
\label{rf6-sec:BI3}

\textit{This section is adapted from the executive summary of a Snowmass white paper solicited by the RF6 conveners and edited by Philip Harris, Philip Schuster, and Jure Zupan~\cite{Harris:2022vnx}.} 

\input{big-idea3}

%% file: experiments.tex
\begin{figure}[h]
  \centering  
  \includegraphics[width=0.97\textwidth]{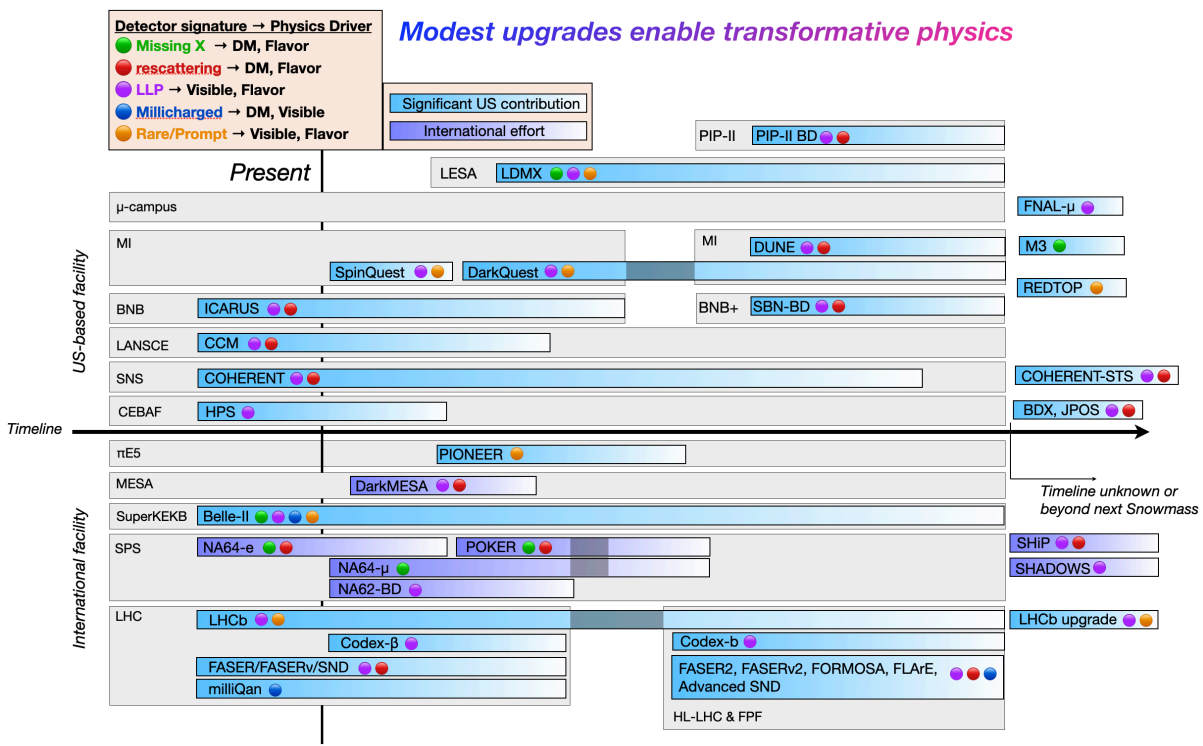}
  \vspace{-3em}
  \caption{Summary of accelerator facilities, experiments, and detector signatures~\cite{Ilten:2022lfq}.\label{rf6-fig:exp-sum}}
\end{figure}

Searches for dark-sector particles in the GeV mass range and below at particle accelerators are a highly-motivated physics opportunity in the next decade. Reference~\cite{Ilten:2022lfq} summarizes and characterizes experiments and facilities for accelerator-based dark-sector searches. The physics drivers are characterized into three main thrusts and those are described in companion reports: thermal DM~\cite{Krnjaic:2022ozp}, visible dark portals~\cite{Batell:2022dpx}, and new flavors and rich dark sectors~\cite{Harris:2022vnx}. Motivated by these physics drivers, \cite{Ilten:2022lfq} enumerates a number of experimental initiatives, describing them and characterizing them by their types of experimental signatures: long-lived particles (LLP), DM rescattering, millicharged particles, missing $X$, and rare prompt decays. Given the large interest in this physics and the number of current and proposed experiments, \cite{Ilten:2022lfq} provides a summary, in one central place and including brief descriptions, of these experiments and facilities. It also provides references to more detailed studies where the reader can find more information. This information is summarized compactly in Fig.~\ref{rf6-fig:exp-sum} based on the beam facility producing the dark-sector particles and the types of detector signatures proposed at those facilities along a schematic timeline.
We note that some potentially exciting but less well developed ideas are not included in \cite{Ilten:2022lfq}, {\em e.g.}, the ambitious Advanced Muon Facility (AMF) proposal would exploit the full potential of the PIP-II accelerator and might provide a superior proton beam dump facility to PIP2-BD and M3~\cite{RF5report}.

%% file: big-idea1.tex
The existence of DM is Nature's sharpest evidence that the SM of particle physics is incomplete. While astrophysical evidence for DM has mounted steadily over the past 8 decades, with increasingly precise measurements confirming the effects of DM from galactic and cluster scales to the primordial early Universe, its particle nature remains elusive.  Identifying the fundamental constituents of dark matter, how these came to dominate the matter density of the Universe, and how they connect to the well-understood physics of ordinary matter are arguably the greatest questions in fundamental physics today.

The space of possible DM masses and properties is vast: the range of viable masses for individual DM constituents spans roughly 50 orders of magnitude; to date its only observed interaction is through gravity; particle properties of DM have not been measured but bulk properties, including its cosmological mass density, inform and motivate models for the DM constituents.   
Indeed, the observed DM density has long served as a goalpost for understanding plausible models of DM, and a hint that suggests DM has microscopic interactions with ordinary matter that are stronger than gravity.  Early thermal equilibrium of DM and familiar matter, followed by freeze-out of the DM as the Universe cools, offers one simple explanation for the origin of its observed abundance.  This freeze-out mechanism is exemplified by the WIMP paradigm, which has long been the focus of terrestrial searches for DM. 

The last decade has seen a tremendous growth of theoretical and experimental interest in DM whose constituents are comparable in mass to electrons or protons, so-called light DM.  This framework simply generalizes the WIMP paradigm to lower masses.  Light DM maintains the simplicity of thermal freeze-out as an origin for DM, as well as the close structural resemblance of the DM sector to the SM, yet poses different experimental challenges and opportunites.  Models of light DM rely for freeze-out on light force-carriers with parametrically weak SM couplings.  As a typical example, a new $U(1)$ gauge boson (dark photon) can mix with the SM photon at the $\sim 10^{-3}$ to $10^{-6}$ level due to radiative effects---a degree of mixing compatible with thermal freeze-out for MeV-to-GeV DM.  These interactions are too weak to be detectable in high-$p_T$ DM searches at high-energy colliders, and the lighter DM particles carry too little kinetic energy to be seen in traditional direct detection.  

\begin{center}
\textbf{Accelerator Production of Light Dark Matter}
\end{center}
\vspace{-1em}
In response to these challenges, laboratory production of light DM by intensity-frontier experiments---including dedicated fixed-target experiments, small forward detectors, and flavor factories---has emerged as an essential strategy for exploring light DM.  These experiments are optimized for intensity, instrumentation precision, and/or background rejection rather than energy reach.  

\begin{figure}[t]
    \begin{center}
    \includegraphics[width=0.9\hsize]{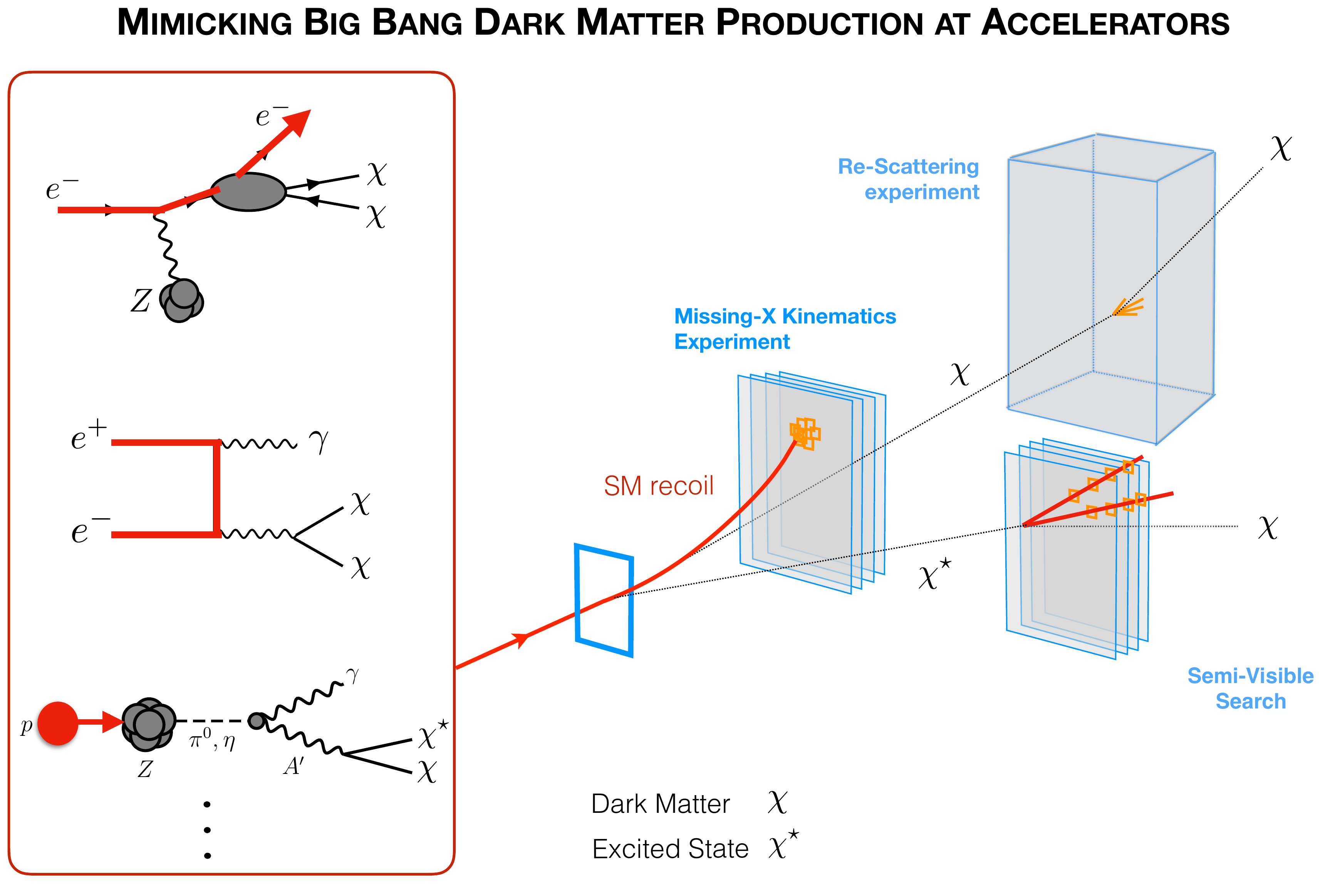}
    \end{center}
    \caption{From \cite{Krnjaic:2022ozp}: Illustration of representative DM production mechanisms (left) and (right) the concepts for detecting DM production via, clockwise from left, missing X, re-scattering, and semi-visible detection strategies. 
    \label{fig:strategyCartoon}}
\end{figure}

Accelerator-based searches for DM exploit several different production mechanisms, including bremsstrahlung-like DM production off beam leptons or protons, meson decays that include DM in the final state, $e^+e^-$ annihilation, and Drell-Yan production.  The search strategies can be grouped into three broad categories.  \emph{Missing energy, momementum, or mass} searches use the kinematics of visible particles recoiling from a DM production event, together with vetoes on SM reaction products, to identify DM production events.  \emph{Re-scattering} experiments search for DM and/or millicharged particles through their subsequent scattering in a detector forward of a fixed-target beam-dump or collider interaction point.  \emph{Semi-visible searches} leverage the possibility of metastable resonances in the dark sector, which can be motivated by specific models of DM cosmology and in many cases decay into a combination of DM and visible SM particles.  These strategies are summarized visually in Fig.~\ref{fig:strategyCartoon} and the program has been reviewed in  Refs. \cite{Essig:2013lka,Battaglieri:2017aum}.

\begin{figure}[t!]
    \begin{center}
    \includegraphics[width=0.9\hsize]{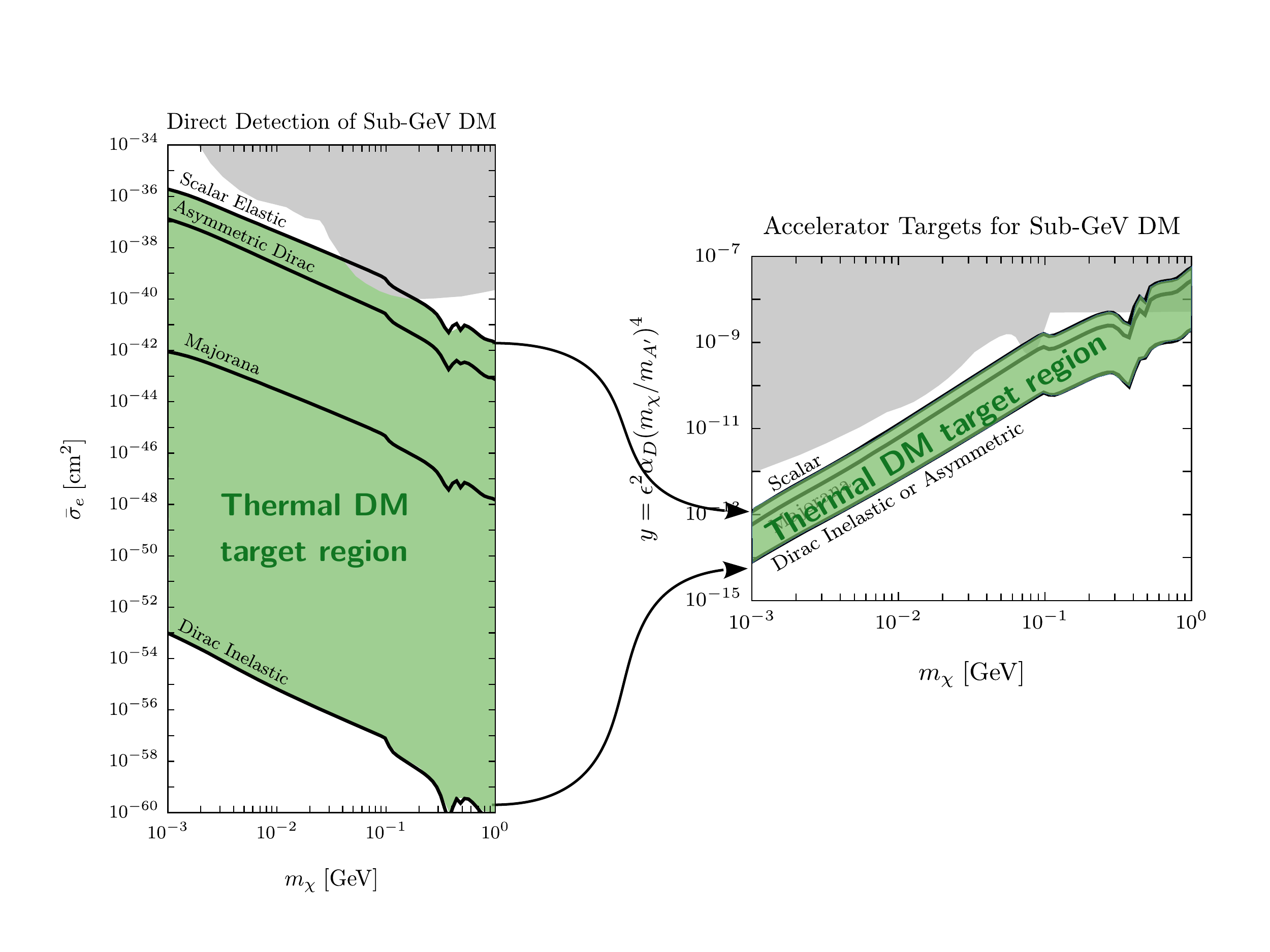}
    \end{center}
    \caption{
    From \cite{Krnjaic:2022ozp}: 
    Comparison of sub-GeV DM thermal production targets for dark-photon-mediated models in the direct detection plane in terms of the electron cross section (left) and on the accelerator plane in terms
    of the variable $y$ (right). Since 
    accelerator production mimics the relativistic kinematics of the early universe, the corresponding signal strength is never suppressed by velocity, spin,
    or small degrees of inelasticity, so the 
    targets are closer to experimentally accessible regions of parameter space. 
    Note, however, that direct detection sensitivity has a complementary
    enhancement for DM candidates with Coulombic
    interactions, which are enhanced at low velocity.
    \label{fig:collapsing}}
\end{figure}

Accelerator-based light DM searches are highly complementary to another promising avenue for discovery of light DM: low-threshold direct detection (discussed in \cite{CF1_Essig:2022dfa}).  Both approaches are essential to a strong light DM search program.  
There are key differences in what properties they probe---accelerator-based experiments directly characterize the particle properties of produced DM, while direct detection explores a combination of these properties with their cosmological abundance. They also probe DM in vastly different kinematic regimes: whereas direct detection probes very non-relativistic scattering, accelerators explore relativistic DM production.  This kinematic difference translates into complementary discovery potential: low-threshold direct detection is a uniquely powerful probe for Coulomb-like interactions with enhancements at low velocities, including models of freeze-in through a light mediator.  Accelerators are optimal for discovery of DM whose interactions are suppressed at low velocities, including thermal freeze-out through a dark photon with generic spin and mass structure.  
As shown in Fig.~\ref{fig:collapsing}, depending on the Lorentz structure 
of the dark-visible interaction, the non-relativistic direct detection cross section can be suppressed
by many orders of magnitude while relativistic accelerator production is not suppressed by such variations.
Still other thermal (and non-thermal) models, including elastically interacting scalar DM, can be observed efficiently by both approaches, allowing exciting opportunities to characterize any observed signal.

\begin{figure}[p!]
\begin{center}
\includegraphics[width=0.9\textwidth]{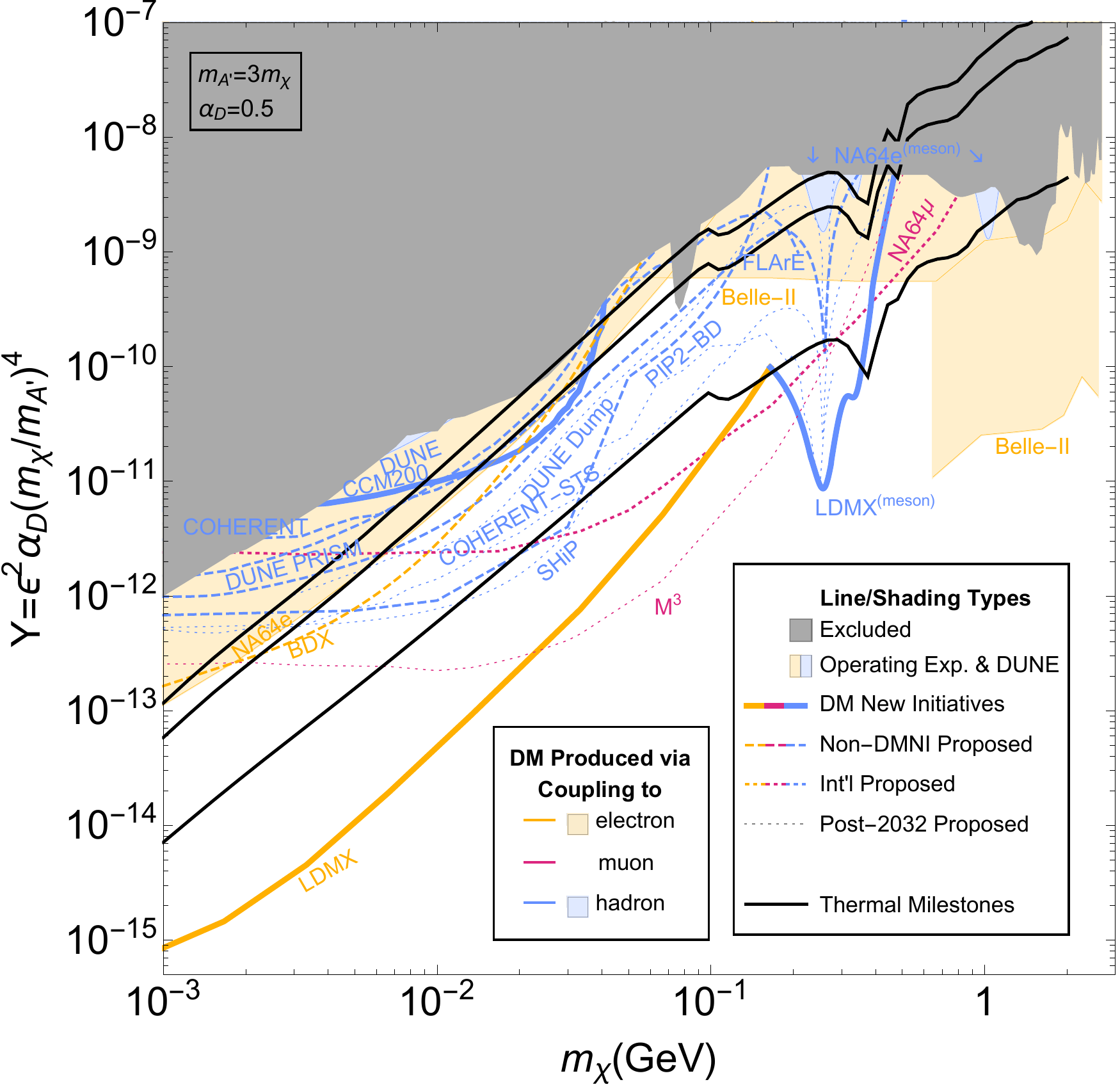}
\end{center}
\caption{
Dark matter production at accelerators~\cite{Krnjaic:2022ozp}: 
Thermal milestones for the kinetically mixed dark-photon model, a key light DM benchmark (shown as black solid lines), along with exclusions from past experiments (gray shaded regions), projected sensitivities of future projects that are operating or have
secured full funding (colored shaded regions), and the experiments partially funded by the Dark Matter New Initiatives (DMNI) program (solid colored lines).  Other proposed experiments that can be realized within a decade are shown as 
long-dashed colored lines if they are based in the US and/or have strong US leadership, or as short-dashed colored lines if they are primarily international efforts. Proposed experiments that are farther into the future are shown as thin dotted lines.  As can be seen, the combination of operating and DMNI experiments can comprehensively explore the thermal DM targets, considered a major milestone of dark-sector physics. 
Each line/region is color-coded according to which SM fermion coupling is employed to produce the DM. In variations of this model, some species may have suppressed couplings, making lines of different colors complementary in this extended model space. \label{rf6-fig:phys-BI1}}
\end{figure}

\begin{center}
\textbf{Science Opportunities and The Road Ahead}
\end{center}
\vspace{-1em}
In the past decade, a key goal of the light DM search effort has been broadly exploring DM models in the MeV-to-GeV mass range. The simplest, and most WIMP-like, viable mechanism for light DM thermal freeze-out is annihilation to SM particles via an $s$-channel dark photon.  This model has therefore emerged as a key benchmark model.  Because DM production at (semi-)relativistic kinematics drives both the dynamics of freeze-out and DM production at accelerators, the range of freeze-out interaction strengths (often parametrized by a dimensionless parameter $y$ related to the effective Fermi scale of the interaction) compatible with this mechanism is narrow, spanning a factor of $\sim 30$ at a given DM mass (black diagonal lines in Fig.~\ref{rf6-fig:phys-BI1}). 

This milestone was identified as a high-priority goal for the accelerator-based program at the DMNI BRN workshop \cite{BRN} and in the subsequent summary 
\href{https://science.osti.gov/-/media/hep/pdf/Reports/Dark_Matter_New_Initiatives_rpt.pdf}{report}.  
Following this, a competitive DMNI process by DOE HEP selected two intensity-frontier projects to support, CCM200 and LDMX, to explore this milestone with different timescales and complementary sensitivity.  CCM200, a proton beam re-scattering experiment at Los Alamos' LANSCE, was completed and commissioned in 2021 and is now operating.  LDMX, a missing momentum experiment at SLAC's LESA electron beamline, received pre-project funds, awaits construction funding, and could begin operation in FY26. CCM200 expands sensitivity to hadronic DM couplings, while LDMX will explore all thermal DM milestones below $\sim 0.5$ GeV, complementing the sensitivity of Belle-II to $\sim$GeV mass thermal DM milestones. These experiments' sensitivity projections 
are illustrated in Fig.~\ref{rf6-fig:phys-BI1}.
The coverage of the thermal milestones by these experiments is robust to many important model dependencies, such as varying dark-sector couplings and the DM to dark-photon mass ratio (excepting a fine-tuned resonance-enhanced region).

Figure~\ref{rf6-fig:phys-BI1} also highlights the DM-search capabilities of many other experimental concepts outside the DMNI scope.  
The breadth of ideas within this program is valuable for several reasons.  The use of multiple complementary techniques will assure a robust program, and in the case of discovery the ability to measure dark-sector masses and interaction strengths. 
Multiple complementary experiments are also important to probe generalizations of thermal freeze-out. Some of these, such as those where a mediator does not couple to electrons but preferentially to $\mu$ and/or $\tau$ leptons or baryons,
motivate a continuing push in missing-$X$ and re-scattering experiments to improve sensitivity to muon and hadron coupled DM production (echoing ``Thrust 1'' of the accelerator Priority Research Direction in \cite{BRN}).  Others, including
models where meta-stable particles in the dark sector play important roles in DM cosmology and enable new discovery techniques, and neutrino portal annihilation with qualitatively different experimental signals, motivate searches for semi-visible DM signals and DM-motivated visible signals (echoing ``Thrust 2'' from \cite{BRN}). We note that DMNI has not yet funded any experiments optimized for (semi)visible dark sector searches, but the next generation of small-project proposals could cover substantial parameter space with strong DM motivation, again complementing Belle II and LHCb capabilities. 

These motivations and corresponding experimental opportunities are 
examined in more detail in \cite{Krnjaic:2022ozp}. 
Related techniques can also advance the detection of millicharged particles, which present a distinctive detector signature and could make up a small fraction of the DM.  Finally, theory has a key role to play in defining the future of the light DM search program---both by continuing to explore the space of light DM models and through theory-experiment collaborations, which have played an important role in the development of many of the concepts and analyses considered here.

The field of concepts in intensity-frontier experiments searching for DM has grown tremendously in the last decade, in response to the tremendous untapped discovery opportunities that it presents. Most of these concepts are low-cost, based on either analyses of multi-purpose experimental data or small experiments that leverage existing accelerator infrastructure and detector technologies.  In the next decade, the realization of 
this opportunity through funding for dark-sector searches, completion of the DMNI-supported program, and the selection of complementary concepts in subsequent round(s) of DMNI
will shed a clear light on the possibility of low-mass particle DM and other light new physics.

%% file: big-idea2.tex
The paradigm of a dark sector comprised of new SM singlet dark particles feebly coupled to ordinary matter through a portal interaction is motivated on a variety of grounds. Dark sectors can resolve some of the outstanding mysteries in particle physics, including the DM puzzle, the dynamics underlying neutrino masses, baryogenesis, the hierarchy problem, and the strong-CP problem. As discussed in the executive summary of this report, dark-sector particles can interact with the SM only through a limited set of (renormalizable or dimension-5) portal operators.
From a bottom-up perspective, portals provide a systematic effective field-theory-based scheme for investigations of new light physics with very weak interactions. The dark-sector framework has proven to be a versatile playground for exploring potential new physics explanations of an array of experimental anomalies. 
Dark-sector research has bloomed worldwide over the past decade with the development of creative theoretical models, the conception of novel phenomenological strategies, and the proposal and implementation of innovative searches and novel experiments. 

Reference~\cite{Batell:2022dpx} covers the physics of {\bf minimal portal interactions}, including the renormalizable vector, Higgs, and neutrino portals, as well as minimal dimension-5 axion-like-particle (ALP) portals with photon or gluon couplings. 
The focus is on minimal extensions of the SM featuring a single new light mediator particle coupled through one of these portals (see next section for non-minimal dark-sector models). This implies that both the production of the mediator at accelerator experiments and its visible decay to SM particles occur due to the portal interaction. 
Reference~\cite{Batell:2022dpx} examines in detail the rich variety of exciting experimental opportunities to investigate the structure of the dark sector by producing and detecting such unstable mediator particles (see also the red region in Fig.~\ref{rf6-fig:flag}). 
It reviews the current status and future prospects for exploring the minimal portals and highlights the connections with some of the big open questions in fundamental physics.

The detailed properties of the mediator, including its mass, spin, and pattern of couplings to the visible sector, are of great interest from both theoretical and phenomenological perspectives. From the theory side, the gauge symmetries and field content of the SM impose tight constraints on the possible nature of the mediator and its couplings. On the other hand, these properties determine, to a significant extent, the possible phenomenological avenues that can be pursued to probe the mediator. In this light, the renormalizable vector, Higgs, and neutrino portals warrant special attention owing to their uniqueness and economy. These portals offer minimal ways to link gauge singlet scalar, fermion, or vector fields to the SM with feeble couplings at low energies. 
Beyond these three portals, the mediation between the visible and dark sectors can occur through higher-dimension portals. 
A well motivated and often studied case is a light ALP, with {\em e.g.}, couplings to photons or gluons through dimension-5 operators, whose mass is protected by a shift symmetry.

Dark sectors are being pursued on multiple experimental fronts with a diverse set of search techniques. Electron and proton beam fixed-target experiments 
with sensitive detectors covering ${\cal O}$(meter -- kilometer) baselines 
provide excellent reach at low dark-particle masses over a broad range of couplings.  Medium energy $e^+ e^-$ colliders/meson factories provide powerful sensitivity for moderate couplings both at low and intermediate masses. Precision studies of meson and lepton decays, including those at pion, kaon, $\eta^{(\prime)}$, and muon facilities, offer interesting and in some cases unique coverage at low masses and small couplings. 
A diverse collection of existing and planned experiments at the LHC, including new LLP detectors, will be able to probe extensive regions of parameter space in a variety of dark-sector models. 
Collectively, these experiments will utilize a wide array of search strategies, including bump-hunt searches for promptly decaying resonances, displaced vertex searches for dark particles with moderate lifetimes, searches for long-lived particle decays to visible final states, and missing momentum searches in both collisions and rare decays.

The minimal portals feature prominently in a variety of proposed solutions to the big questions in fundamental physics. 
These include a variety of motivated dark matter scenarios with novel cosmology and phenomenology. One generic example is secluded dark matter, in which heavier dark matter is thermally produced in the early universe via its annihilation to lighter mediator particles. Viable secluded DM models can be realized in any of the minimal portals. 
The requirement of thermalization in secluded scenarios imposes a lower bound on the portal coupling, offering an interesting target for high intensity probes. 
A variety of other interesting DM scenarios in which the mediator is the lightest dark-sector state have been proposed, many of which can be correlated with specific regions of parameter space within the minimal portal models. The minimal portals may also play an important role in solutions to puzzles motivated by naturalness considerations. In particular, the Higgs portal is a critical ingredient in the relaxion solution to the hierarchy problem, while the vector portal is expected on general grounds and may have important consequences in the mirror Twin Higgs model, which tackles the little hierarchy problem. 
The ALP portal, and in particular the ALP-gluon interaction, is motivated by its connection to the strong-CP problem. The neutrino portal is likely to offer an explanation of the light SM neutrino masses and may also give rise to low-scale leptogenesis mechanisms. Furthermore, a light scalar interacting via the Higgs portal may also serve as the inflaton. The Snowmass white paper \cite{Batell:2022dpx} spotlights the myriad connections between the minimal portal models and the potential answers to the big questions in particle physics and cosmology. 

The theoretical ideas and experimental approaches featured in \cite{Batell:2022dpx} have important synergies and complementarity with other efforts across the rare and precision, energy, cosmic, and neutrino frontiers. At the energy frontier, the LHC and future high-energy colliders will be able to probe heavier mediators with larger couplings. Distinct experimental opportunities are also available at the energy frontier, including {\em e.g.}, exotic Higgs decays and precision measurements of Higgs couplings and electroweak observables. At the cosmic frontier, a suite of new direct detection experiments will directly search for halo DM through its scattering, while an array of astrophysical observations can indirectly search for signatures of DM annihilation. In particular, these direct and indirect searches provide sensitivity to DM that is heavier than the mediator and are therefore highly complementary to the direct searches for the visibly decaying mediators at intensity frontier experiments. There is also an exciting interplay with activities in the neutrino frontier. Dark-sector mediators can be sought at accelerator- and reactor-based neutrino experiments and have also been invoked in a variety of potential BSM explanations for various experimental anomalies in the neutrino sector ({\em e.g.}, the MiniBooNE low energy excess of electron like events).

\begin{figure}[p!]
\begin{center}
\includegraphics[width=0.9\textwidth]{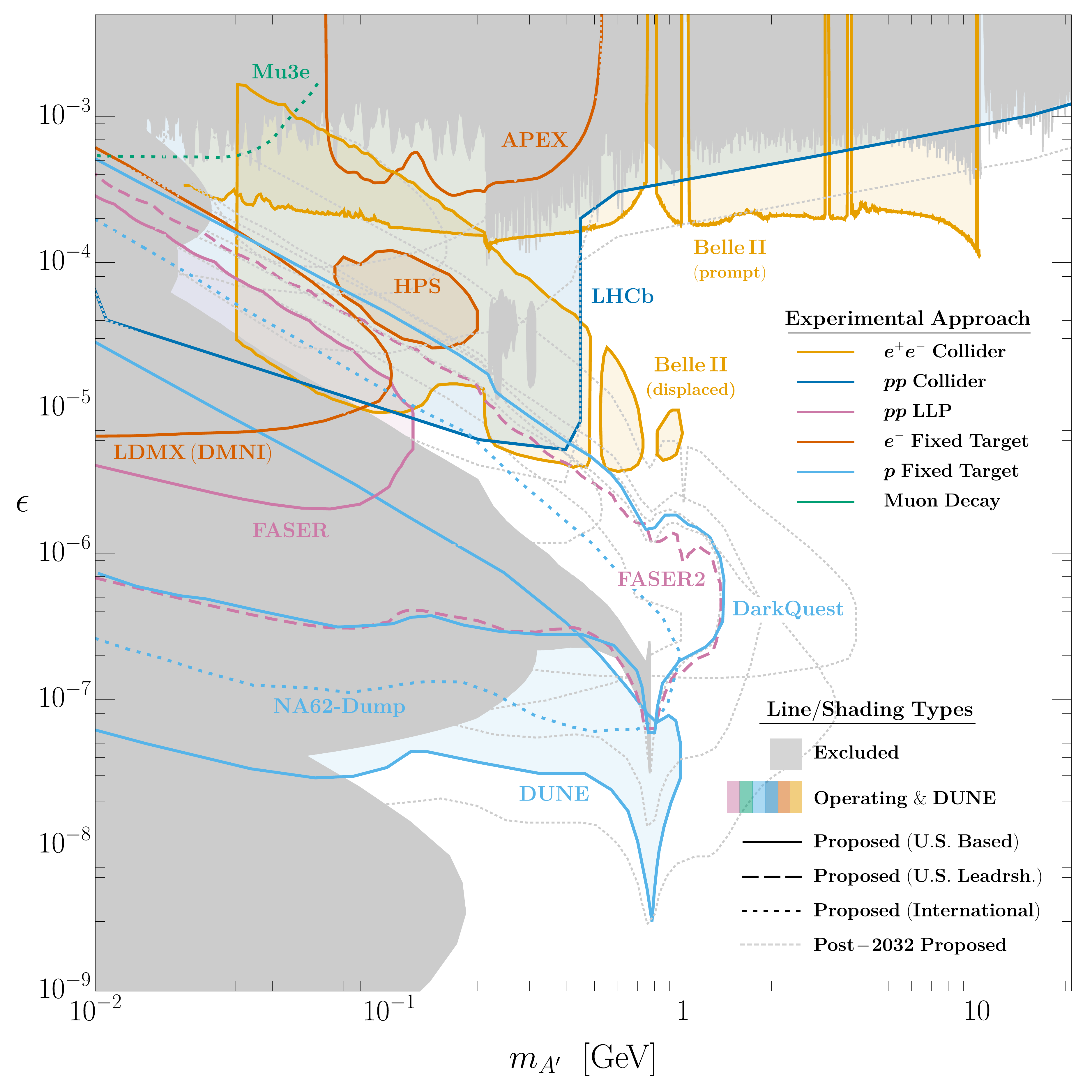}
\end{center}
\caption{ 
Visible dark-photon parameter space (compatible with secluded thermal DM)~\cite{Batell:2022dpx}: 
Near-term and future opportunities to search for visibly decaying dark photons interacting through the vector portal displayed in the dark photon mass $(m_{A'})$ -- kinetic mixing  $(\epsilon)$ parameter space. 
Constraints from past experiments (gray shaded regions) and projected sensitivities from operating and fully funded experiments and DUNE (colored shaded regions), proposed near-term (pre-2032) experiments based in the US including DMNI supported experiments (solid colored lines), proposed near-term (pre-2032) experiments based internationally and having significant US leadership (dashed colored lines), proposed near-term (pre-2032) international projects (dotted colored lines), and proposed future (post-2032) experiments (dotted gray lines) are shown. 
Line coloring indicates the key experimental approach used ($e^+ e^-$ collider, $pp$ collider, LHC LLP detector, electron fixed target, proton fixed target, muon decay), highlighting one aspect of the complementarity between different facilities/experiments. Collectively, these experiments are poised to cover large regions of open dark photon parameter space. 
}
\label{rf6-fig:phys-BI2-aprime}
\end{figure}

\begin{figure}[p!]
\begin{center}
\includegraphics[width=0.9\textwidth]{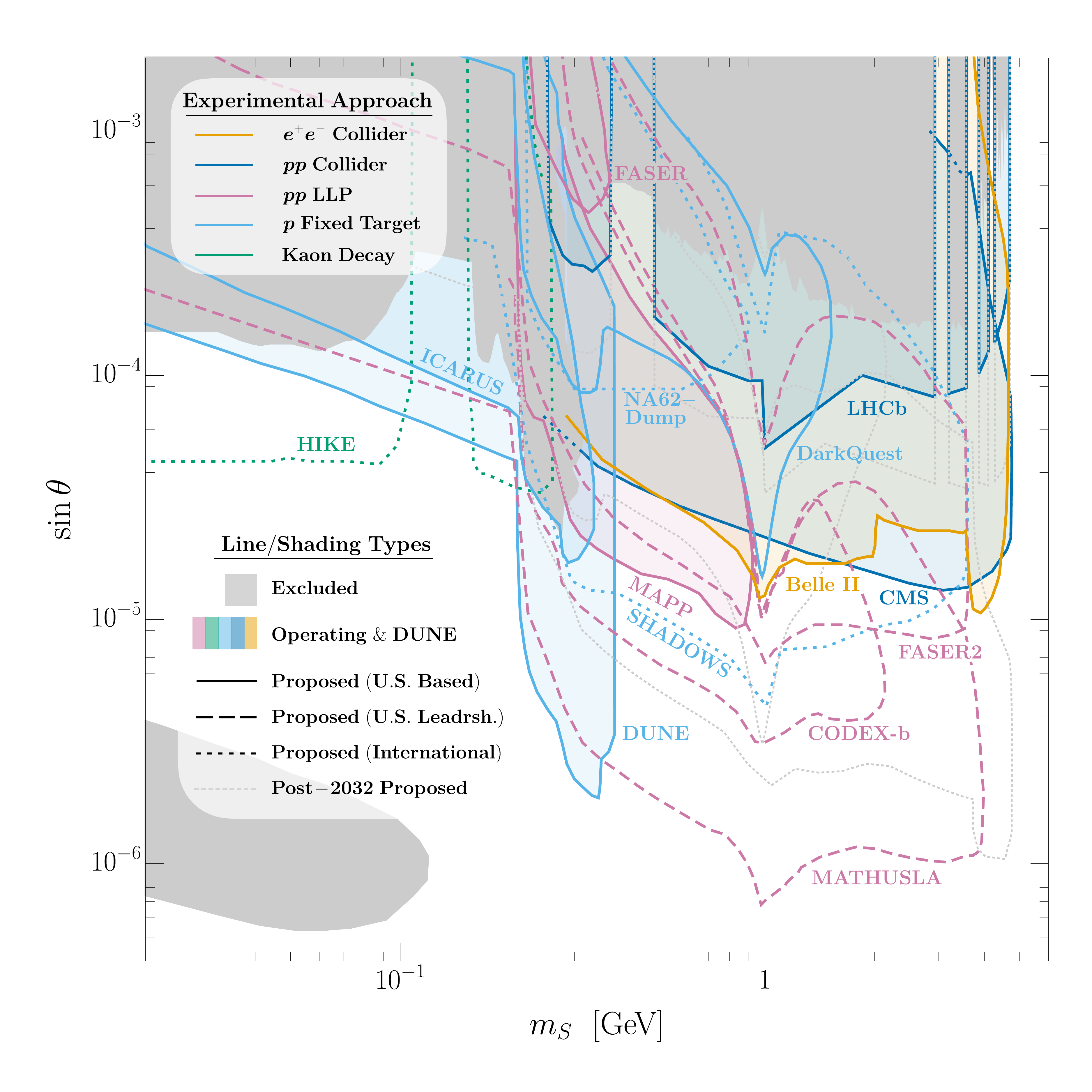}
\end{center}
\caption{
Visible dark-scalar parameter space (compatible with secluded thermal DM) ~\cite{Batell:2022dpx}: 
Near-term and future opportunities to search for visibly decaying dark scalars interacting through the Higgs portal displayed in the scalar mass $(m_{S})$ -- mixing angle $(\sin\theta)$ parameter space. Constraints from past experiments (gray shaded regions) and projected sensitivities from operating and fully funded experiments and DUNE (colored shaded regions), proposed near-term (pre-2032) experiments US based (solid colored lines), proposed near-term (pre-2032) experiments based internationally and having significant US leadership (dashed colored lines), proposed near-term (pre-2032) international projects (dotted colored lines), and proposed future (post-2032) experiments (dotted gray lines) are shown. 
Line coloring indicates the key experimental approach used ($e^+ e^-$ collider, $pp$ collider, LHC LLP detector, proton fixed target, kaon decay), highlighting one aspect of the complementarity between different facilities/experiments. Collectively, these experiments are poised to cover large regions of open dark scalar parameter space.
}
\label{rf6-fig:phys-BI2-higgs}
\end{figure}

As illustrative examples, Figs.~\ref{rf6-fig:phys-BI2-aprime} and \ref{rf6-fig:phys-BI2-higgs} present the near-term and future opportunities to probe the minimal vector portal and Higgs portal models, respectively.
Dark photons and Higgs portal scalars are well-motivated dark-sector benchmarks, may serve as  mediators to dark matter, and appear in a variety of UV models addressing big open questions in particle physics. 
A combination of operating, fully or partially funded, and proposed near-term and future experiments will be able to search for dark photons and dark Higgs bosons over a broad range of currently unconstrained parameter space. As these examples illustrate, and as is highlighted in \cite{Batell:2022dpx} with specific case studies, there is great opportunity to explore vast uncharted parameter space and investigate the structure of the dark sector during the next decade and beyond.

The dark-sector science program has developed significantly during the last decade. 
Milestones in this trajectory can be seen in past community studies, including the Dark Sectors 2016 Workshop~\cite{Alexander:2016aln}, the US Cosmic Visions New Ideas in Dark Matter 2017~\cite{Battaglieri:2017aum}, and the CERN Physics Beyond Colliders initiative~\cite{Beacham:2019nyx}. Furthermore, the US DOE Office of Science Dark Matter New Initiatives (DMNI) Basic Research Needs (BRN) report~\cite{BRN} has recently highlighted the importance of MeV-GeV scale dark-sector studies as a Priority Research Direction: ``Create and detect dark matter particles below the proton mass and associated forces, leveraging DOE accelerators that produce beams of energetic particles.'' The experiments discussed in \cite{Batell:2022dpx} would uniquely contribute to the {\it{thrust 2}} ``Explore the structure of the dark sector by producing and detecting unstable dark particles'' highlighted in that report.

Realizing the broad objectives of the dark-sector science program, as spotlighted by Secs.~\ref{rf6-sec:BI1}--\ref{rf6-sec:BI3}, necessitates efforts and investments in several primary directions, including harnessing the potential of existing large-scale multi-purpose detectors, supporting dedicated small-scale experiments and facilities with high-intensity beams, and promoting advances in dark-sector theory and fostering the community of dark-sector theorists. Support for these projects and research activities will facilitate a broad and high-impact dark-sector physics program during the next decade, with US scientists and institutions providing key leadership in this effort.

%% file: big-idea3.tex
Dark matter directly  motivates the existence of a dark sector of matter and interactions beyond the Standard Model. To-date, much of the emphasis for experimental work on dark sectors has been anchored to minimal models, often with only a single mediator particle, single DM candidate, and the assumption of flavor universality in the interactions. However, like the Standard Model, dark sectors may have non-minimal structures, either in couplings to the SM, or in the spectra of dark-sector states. Often this leads to a far richer phenomenology and may require new experimental strategies for  achieving optimized sensitivities. Additionally, extensions of minimal dark-sector scenarios can help to resolve important theoretical puzzles and data-driven anomalies. The purpose of \cite{Harris:2022vnx} is to showcase examples where interesting non-minimal dark-sector phenomena can be efficiently searched for and uncovered in high-intensity experiments. These examples are organized into three themes: (1) examples motivated by data-driven anomalies, (2) examples motivated by theoretical puzzles, (3) phenomenological examples motivated by the desire to achieve more
complete coverage of the standard benchmark models beyond the assumption of minimality.

From a theory point of view, dark sectors coupled to the SM by means of the so-called portal operators 
represent a framework that encapsulates a broad range of new physics ideas while respecting the well-measured symmetries of the SM. As such, it is not surprising that the framework of dark-sector portals is widely used to study new physics explanations for a range of data-driven anomalies, such as those in the measurements of 
$(g-2)_\mu$ and the quark-flavor sector~\cite{HFLAV:2019otj}, among others. 
 
\begin{figure}[p!]
    \centering
     \includegraphics[width=0.9\textwidth]{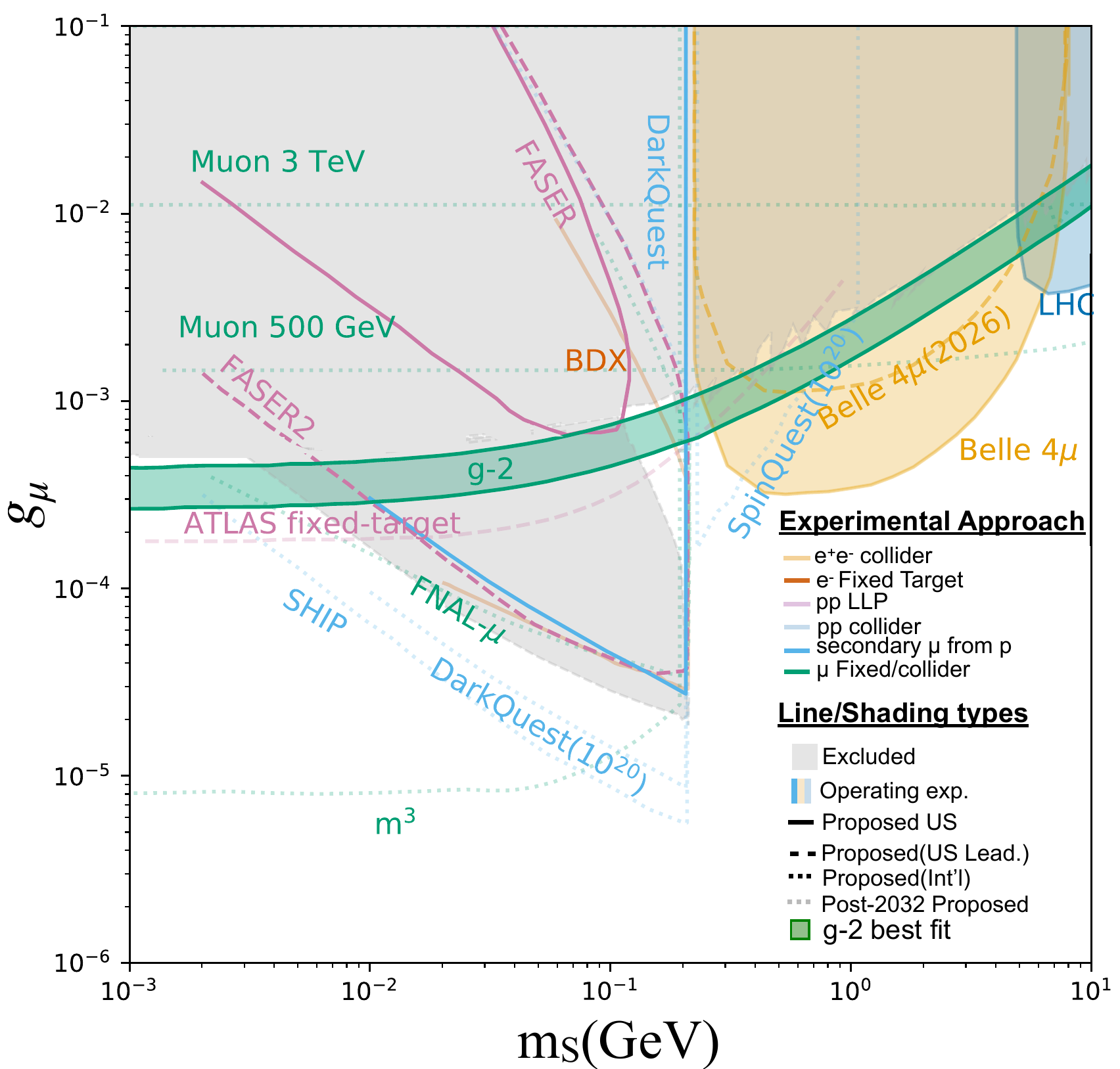}
    \caption{Muon-philic scalar model~\cite{Harris:2022vnx}: The parameter space that can explain the $(g-2)_\mu$ anomaly is shown as a green band and represents the main experimental target of this model.
    Constraints from past experiments (gray shaded regions) and projected sensitivities from operating and fully funded experiments (colored shaded regions), proposed near-term (pre-2032)
experiments based in the US (solid colored lines) proposed near-term (pre-2032)
experiments based internationally and having significant US leadership (dashed colored lines), proposed near term
(pre-2032) international projects (dotted colored lines), and proposed future (post-2032) experiments (dotted gray lines). Line coloring indicates the key experimental approach used.
   } 
   \label{rf6-fig:phys-BI3-g-2}
 \end{figure}

Among these, $(g-2)_\mu$ is the  longest-standing data-driven anomaly. 
Reference~\cite{Harris:2022vnx} features several new physics models that can explain the discrepancy between the measurements and the consensus SM prediction. In the case of the $(g-2)_\mu$ anomaly the program of dark-sector experiments has already had considerable impact, ruling out the simplest new physics ideas that were put forward to address the anomaly. The parameter space that remains involves non-minimal flavor-dependent interactions (see Fig.~\ref{rf6-fig:phys-BI3-g-2}). High-intensity dark-sector experiments will be able to probe most of the remaining explanations in the coming years, either ruling out the new physics explanations or making discoveries.

\begin{figure}[p!]
\centering
\includegraphics[width=0.9\textwidth]{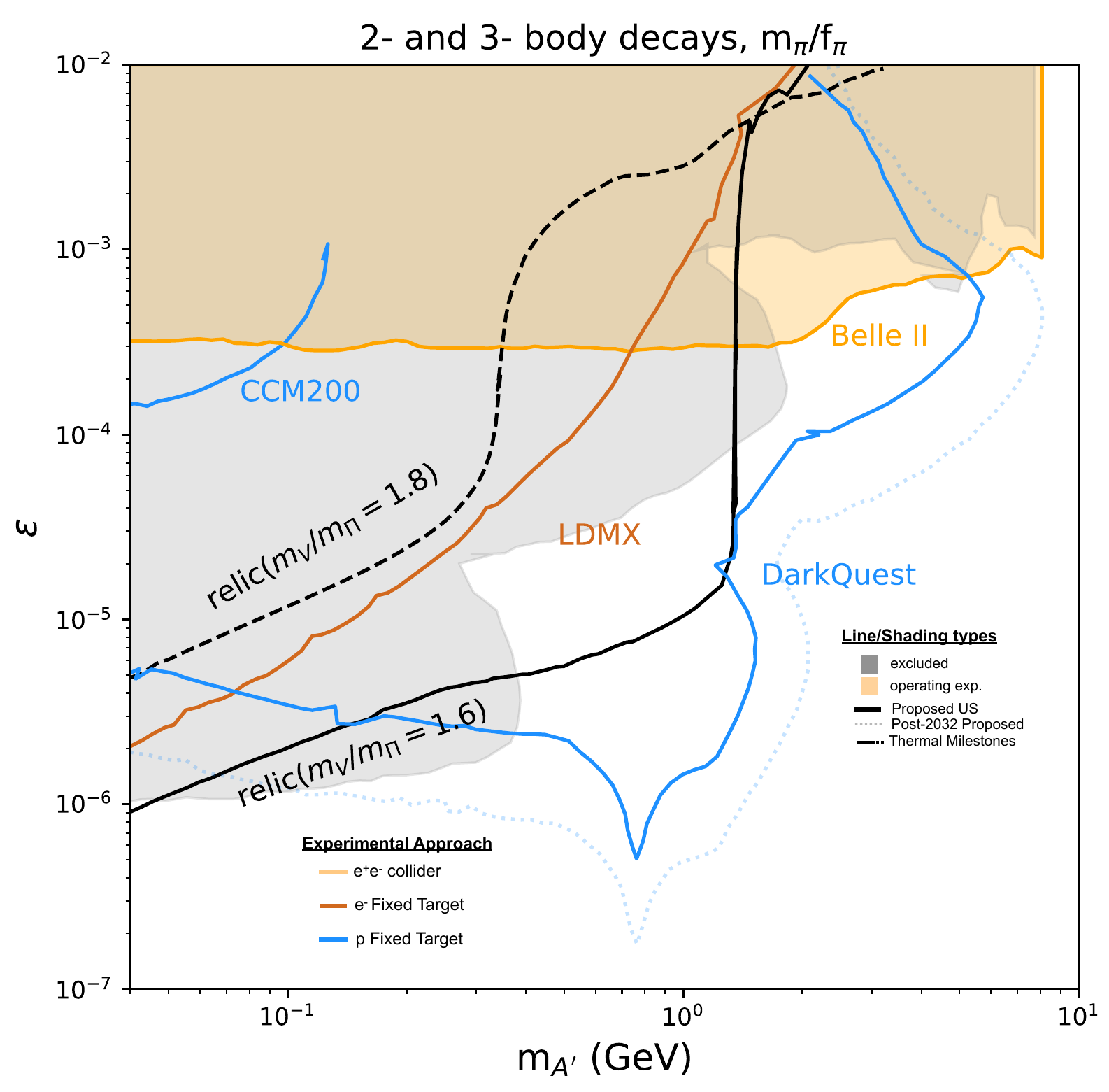}
\caption{An example of the SIMP parameter space for a QCD-like strongly coupled dark sector gauged under a new $U(1)$ with a dark photon with interactions to the SM through the mixing parameter $\epsilon$ (see \cite{Harris:2022vnx} for details about the dark-sector model parameters).
 Constraints from past experiments (gray shaded regions) and projected sensitivities from operating experiments (colored shaded regions), along with projected sensitivities for proposed near-term experiments (dashed colored lines). 
 Dark pions constitute all of the observed DM abundance on solid (dashed) black
  contours, while DM is overabundant below these lines. 
\label{rf6-fig:phys-BI3-simp}}
\end{figure}

The range of more complete models that include the most common minimal dark-sector benchmarks is rather large, see, {\em e.g.}, Ref.~\cite{Battaglieri:2017aum}.  
This is part of the reason the minimal models have so far received much more attention of the community. A common element of the more complete models concerns the origin of the dark-sector mass scale, with the models split  into two categories---the weak and the strong coupling regimes. 
Reference~\cite{Harris:2022vnx} gives a representative example from each category, namely inelastic dark matter (iDM), and the strongly interacting massive particles (SIMPs). These are two interesting models that can predict a thermal DM candidate with the measured relic abundance. Many minimal dark-sector DM benchmarks include a massive vector mediator and a scalar or fermion DM candidate. In UV completions of these scenarios, the dark Higgs that gives mass to the vector can also split the DM multiplet into a pair of nearly degenerate states, the lightest of which is stable. This is what happens in iDM models. Similar phenomenology can be obtained in the second category of models (SIMPs), in which the vector mediator mass arises from strong dynamics. Strong dynamics naturally results in dark-sector mesons, and thus also predicts a wide range of new phenomena with displaced decays of the mesons to final states with SM leptons.  For both iDM and SIMPs, regions of parameter space with viable predictions for the relic density exist. However, the dynamics responsible for the DM relic abundance changes with respect to the minimal portal benchmark models. SIMPs models utilize 3$\rightarrow$2 annihilation processes to deplete the DM abundance in the early universe. iDM models utilize the cohannihilation of DM with a heavier state. 
Experiments in the next 10 years will be able to probe a significant portion of the allowed parameter space that would lead to a viable DM relic density; see Fig.~\ref{rf6-fig:phys-BI3-simp}.

The dark-sector framework includes generalizations of the long sought after QCD axion, $a$, parametrizing it as part of a more general class of pseudoscalar dark-sector particles, referred to as axion-like particles (ALPs). The high-intensity dark-sector experimental program will  probe a large region of uncharted parameter space for all ALP couplings to the SM particles, including flavor-violating couplings to quarks and leptons, for a broad range of masses. In this way, one probes also the non-minimal QCD axion models, in many cases testing complementary parameter space to the searches based on axion couplings to photons, with the reach above astrophysics constraints. 
Reference~\cite{Harris:2022vnx} highlights this point for the case of a flavor-violating QCD axion model, where searches for $s\to d a$ and $\mu \to e a$ transitions can probe the parameter range in which the QCD axion is a viable cold DM candidate.

The above examples illustrate the robustness of the growing dark-sector experimental effort, both the broadness of the different searches as well as the precision, probing theoretically well-motivated parameter space.